\documentclass[12pt,preprint]{aastex}
\def\mr{Maggiore-Riotto}
\def\ca{Corasaniti-Achitouv}
\def\st{Sheth-Tormen}
\def\tn{Tinker {\it et al.}}
\shortauthors{Ryu \& Lee}
\begin{document}
\title{The Diffusion Coefficient of the Splashback Mass Function as a Probe of Cosmology}
\author{Suho Ryu and Jounghun Lee}
\affil{Astronomy program, Department of Physics and Astronomy,
Seoul National University, Seoul  08826, Republic of Korea \\
\email{shryu@astro.snu.ac.kr, jounghun@astro.snu.ac.kr}}
\begin{abstract}
We present an analytic model for the splashback mass function of dark matter halos, which is parameterized by a single coefficient  
and constructed in the framework of the generalized excursion set theory and the self-similar spherical infall model. 
The value of the single coefficient that quantifies the diffusive nature of the splashback boundary is determined at various redshifts 
by comparing the model with the numerical results from the Erebos N-body simulations for the Planck and the WMAP7 cosmologies. 
Showing that the analytic model with the best-fit coefficient provides excellent matches to the numerical results in the mass range of 
$5\le M/(10^{12}h^{-1}\,M_{\odot})< 10^{3}$,  
we employ the Bayesian and Akaike Information Criterion tests to confirm that our model is most preferred by the numerical results to the previous 
models at redshifts of $0.3\le z\le 3$ for both of the cosmologies. It is also found that the diffusion coefficient decreases almost linearly with redshifts, 
converging to zero at a certain threshold redshift, $z_{c}$, whose value significantly differs between the Planck and WMAP7 cosmologies.  
Our result implies that the splashback mass function of dark matter halos at $z\ge z_{c}$ is well described by a parameter-free analytic 
formula and that $z_{c}$ may have a potential to independently constrain the initial conditions of the universe. 
\end{abstract}
\section{Introduction}\label{sec:intro}

The concept of a splashback radius of a dark matter (DM) halo, defined as the distance from the halo center to the apocenter of the first 
orbit of the newest infalling objects \citep{adh-etal14,mor-etal15} has recently risen up to limelight. 
Unlike the virial radius, it properly includes the flyby subhalos \citep{mor-etal15},  does not suffer from pseudo-evolution \citep{die-etal13}, 
and coincides well with the location on the halo periphery where the stacked density profile exhibits an abrupt decline 
\citep{wan-etal09,wet-etal14,DK14,adh-etal14,die20c}. 
Given these favorable properties, the splashback radius is often dubbed "a physically motivated boundary", which naturally sunders objects 
orbiting in the inner region of a halo from still infalling ones in its outer region \citep{mor-etal15}. 

Very recently, \citet{die20a} found another auspicious aspect of the splashback boundary from a N-body experiment.  
Determining the splashback mass of each DM halo as the total mass of the DM particles enclosed by its splashback radius, 
he constructed an analytic formula for the splashback mass function by employing the excursion set ansatz and 
claimed that it exhibits a substantially higher degree of universality than the virial mass function. 
In the excursion theory, the mass function of DM halos is referred to as being universal, if its multiplicity 
(see Section \ref{sec:review} for a definition) does not have explicit dependence on the redshift and shape of the linear density 
power spectrum \citep{jen-etal01}.   The success of the nearly universal form of the splashback mass function provided by \citet{die20a} 
was demonstrated by its good agreements with the numerical results at all redshifts for the case of the self-similar cosmologies as well as 
for the standard cosmology where the cosmological constant ($\Lambda$) and cold dark matter (CDM) are the most dominant energy and matter 
contents of the universe, respectively, whether the initial conditions are described by the key cosmological parameters from the Planck 
experiment \citep{planck14} or from the Seven-Year Wilkinson Microwave Anisotropy Probes \citep[hereafter, WMAP7][]{wmap7}. 

The excursion set theory describes the growth of the initially overdense regions into DM halos as a random walking process bounded by a 
prescribed absorbing barrier, and evaluates the virial mass function of DM halos by enumerating the random walks that hit for the first time the 
absorbing barrier on a given virial mass scale \citep{bon-etal91}.  The degree of the universality of the virial mass function is contingent upon the shape, 
height and statistical nature of the absorbing barrier, which in turn is determined by the underlying collapse dynamics, background cosmology, and  
halo-identification procedure \citep{CL01,SMT01,MR10a,MR10b}. 
For instance, the absorbing barrier would have a flat shape and redshift-independent constant height, yielding a completely universal mass function, 
if the DM halos are formed via spherically symmetric collapse process and identified without any ambiguity in an Einstein de Sitter universe where 
the matter density parameter equals unity ($\Omega_{\rm m}=1$) \citep{PS74,bon-etal91}.

The departure of the true gravitational collapse from spherical symmetry and the presence of dark energy lead the absorbing barrier 
to have a curved shape with weakly $z$-dependent height \citep{lah-etal91,eke-etal96,KS96,SMT01}, which in consequence slightly decreases 
the degree of the universality of the excursion set mass function of DM halos. 
Moreover, in realistic situations, the virial radius  of a DM halo is susceptible to the pseudo-evolution \citep{die-etal13} as well as to the tidal disturbance 
from the surrounding cosmic web, which implies that considerable amount of ambiguity should be involved in the determination of the virial mass 
\citep{rob-etal09,MR10a,MR10b}. To take into account this ambiguity, the absorbing barrier degrades into a stochastic variable whose standard 
deviation varies with redshift and key cosmological parameters \citep{MR10b,CA11a,CA11b}.  Thus, the degree of the universality 
of the excursion set virial mass function is most closely linked with that of the stochasticity of the absorbing barrier. 

Meanwhile, if the boundary of a halo is set at its splashback radius which does not experience pseudo-evolution and incorporates properly the 
orbits of the flyby objects ejected by the tidal disturbance \citep{mor-etal15}, 
the identification of a DM halo and determination of its mass are expected to suffer much less from ambiguity, rendering the absorbing barrier 
to become less stochastic. The detected higher degree of universality of the splashback mass function in the seminal work of \citet{die20a} can 
be naturally explained by this reduced degree of stochasticity of the absorbing barrier. 
What \citet{die20a} focused on, however, was not providing a physical explanation for the nearly universal form of the spashback mass 
function in terms of the less stochastic absorbing barrier, but finding a practical formula for it characterized by redshift independent multiple parameters. 
In other words, even though it used the excursion set ansatz and turned out to be universal and robust against the variations of the background cosmologies,  
the formula of \citet{die20a} characterized by multiple free parameters is not a physical model but a mere fitting formula.  

Here, we will attempt to find a physical model for the splashback mass function in the generalized excursion set formalism. 
Instead of preoccupying with expressing the splashback mass function in an universal form, we will focus on deriving it from the first principles and reducing 
the number of free parameters that characterize it. The plan of this paper is as follows. 
In Section \ref{sec:review}, we review the excursion set theory for the virial mass function. In Section \ref{sec:absorbing} we present how the slashback 
mass function can be analytically obtained in the generalized excursion set theory. In Section \ref{sec:numerical}, we present the results of numerical 
testing and model selection. In Section \ref{sec:con} we summarize the implication of our findings and discuss its prospect as a cosmological probe as well.

\section{Splashback Mass Function in the Generalized Excursion Set Theory}\label{sec:test}

\subsection{Review of the excursion set mass function theory}\label{sec:review}

The virial mass function of DM halos, $dN(M,z)/d\ln M$, gives the number densities of DM halos whose virial masses fall in a logarithmic 
mass interval of $[\ln M , \ln M+d\ln M]$ at redshift $z$. It is usually inferred from the differential volume fraction, $dF/d\ln M$, occupied by the 
corresponding protohalo sites in the initial density field, $\delta({\bf x})$, smoothed on the scale of $M$  \citep{PS74}: 
\begin{equation}
\label{eqn:massft}
\frac{dN(M, z)}{d\ln M} =  \frac{\bar{\rho}}{M}\Bigg{\vert}\frac{dF(M,z)}{d\ln M}\Bigg{\vert}\, ,
\end{equation}
where the ratio of the mean mass density of the universe to the virial mass, $\bar{\rho}/M$, represents the inverse of the mean volume of the protohalo, 
$\sigma(M,z)$ denotes the rms density fluctuation related to the linear density power spectrum, $P( k,z)$, as 
\begin{eqnarray}
\label{eqn:sig}
\sigma^{2}(M,z)&=&\frac{1}{2\pi^{2}}\int^{\infty}_{0}k^{2}P(k,z)W^{2}[k,M(R_{f})]dk \\ 
W[k, M(R_{f})] &=& \frac{3\left[\sin(kR_{f})  - kR_{f}\cos(kR_{f})\right]}{(kR_{f})^{3}} \, ,
\end{eqnarray}
where the halo mass $M$ is related to the filtering radius $R_{f}$ as $M=4\pi\bar{\rho} R^{3}_{f}/3$ \citep{bon-etal91}.
Throughout this paper, we will exclusively use the CAMB code \citep{camb} for the computation of $P(k,z)$. 

The excursion set mass function theory portrays the evolution of a proto-halo as a random walk process in the $\sigma^{-1}$-$\delta$ configuration 
space, expressing the differential volume fraction in terms of a multiplicity function, $f(\sigma)$, as 
\begin{equation}
\label{eqn:diffvol}
\left\vert\frac{dF}{d\ln M}\right\vert =  \Bigg{\vert}\frac{d\ln\sigma^{-1}}{d\ln M}\Bigg{\vert}f[\sigma(M, z)]\, ,
\end{equation}
where $f(\sigma)$ gives the number of the random walks that are terminated after hitting for the first time the absorbing density barrier, 
$\delta_{c}$, at a given $\sigma^{-1}$. If no parameter of $f(\sigma)$ depends explicitly on $z$ and $\Delta (\ln k, z)$, then the excursion set 
mass function is referred to as being universal. 

For the unrealistic special case that the DM halos form in an Einstein de Sitter universe via the top-hat spherical collapse of the overdense protohalo sites 
in the linear density field smoothed by the sharp $k$-space filter, which is translated into the excursion set jargon, {\it the Markovian random walks bounded 
by a flat (i.e., mass-independent) deterministic absorbing barrier}, \citet{bon-etal91} analytically derived the multiplicity function as 
\begin{equation}
\label{eqn:ps}
f(\sigma) = \sqrt{\frac{2}{\pi}}\frac{\delta_{sc}}{\sigma} \exp\left[-\frac{\delta_{sc}^2}{2\sigma^2}\right]\, .
\end{equation}
Here, $\delta_{sc}=1.686$ represents the value of $\delta_{c}$, called the spherical height of the absorbing barrier, for this special case. It is obvious 
from Equation (\ref{eqn:ps}) that for this special case the virial mass function is universal. 
For the case of a $\Lambda$CDM universe with matter density parameter of $\Omega_{m}<1$, the height of the absorbing barrier $\delta_{c}$ slightly 
deviates from the Einstein-de Sitter value of $\delta_{sc}=1.686$, exhibiting a very weak dependence on redshift, i.e., $\delta_{sc}=\delta_{sc}(z)$ 
\citep{lah-etal91,eke-etal96,KS96}, which in turn marginally decreases the degree of universality of the virial mass function. 

Finding that Equations (\ref{eqn:massft})-(\ref{eqn:ps}) fail in matching at a quantitative level the numerical results from N-body simulations and realizing 
that for the case of a more realistic elliptical collapse dynamics the absorbing barrier must have a curved shape, i.e., $\delta_{c}=\delta_{c}(M)$, 
\citet[][hereafter, ST]{ST99} proposed the following more complicated form of the multiplicity function: 
\begin{equation}
\label{eqn:st}
f(\sigma;A,a,p) = A \sqrt{\frac{2a}{\pi}} \left[ 1 + \left(\frac{\sigma^2}{a\delta_{sc}^2}\right)^p \right] \frac{\delta_{sc}}{\sigma} \exp\left[-\frac{a\delta_{sc}^2}{2\sigma^2}\right]\, , 
\end{equation}
where $A,\ a,\ p$ are three adjustable parameters.  Although the follow-up work provided a theoretical explanation for the curved shape of the absorbing 
barrier used to derive Equation (\ref{eqn:st}) \citep{SMT01}, the \st\ mass function had to resort to N-body simulations for the empirical determination 
of the three parameters at the cost of achieving much better agreements with the numerical results in a wide redshift range . 

\citet{MR10a,MR10b} derived a more complicated but realistic physical model for the cluster mass function, pointing out that in the cluster mass section 
the primary cause of deviation of $f(\sigma)$ from Equation (\ref{eqn:ps}) is not the departure of the real gravitational collapse from the top-hat spherical 
dynamics  but the stochastic nature of the halo identification procedure. 
Basically, they generalized the excursion set theory by treating the absorbing barrier, $\delta_{c}$, as a stochastic variable with average height of 
$\delta_{sc}$ (i.e., $\langle \delta_{c}\rangle = \delta_{sc}=1.686$)\footnote{\citet{MR10a} set $\delta_{sc}$ at the Einstein-de Sitter value of $1.686$, 
arguing that the exact value of $\delta_{sc}$ for the $\Lambda$CDM universe shows only marginal discrepancy from the Einstein de-Sitter value.}, 
and describing the density growth as {\it non-Markovian random walks} (corresponding to using more realistic filters to smooth the density field), and  
succeeded in deriving a purely analytical multiplicity function with the help of the path integral method: 
\begin{eqnarray}
\label{eqn:mr}
f(\sigma ; D_{B}) &\approx& \frac{\delta_{sc}}{\sigma}\sqrt{\frac{2}{\pi(1+D_{B})}} \Biggl\{ \left[1-\frac{\kappa}{(1+D_{B})}\right]
\exp\left[-\frac{\delta_{sc}^2}{2\sigma^2(1+D_{B})}\right]  \nonumber \\
&&+\frac{\kappa}{2(1+D_B)}\Gamma\left[0,\frac{\delta_{sc}^2}{2 \sigma^2(1+D_{B})}\right]\Biggl\}\, .
\end{eqnarray}
Here, $\Gamma$ is the incomplete gamma function, $D_{B}$ is a diffusion coefficient newly introduced by \citet{MR10a} to quantify the degree of the 
stochasticity of the absorbing barrier $\delta_{c}$, and $\kappa$ is a non-Markovian coefficient that quantifies the first order deviation of of the density 
cross correlations on two different mass scales of $M$ and $M^{\prime}$  from its zeroth order Markovian value: 
\begin{equation}
\label{eqn:kappa}
\langle\delta(M)\delta(M^{\prime})\rangle \approx 
\sigma^{2}(M^{\prime})+ \kappa \frac {\sigma^{2}(M^{\prime})\left[\sigma^{2}(M)-\sigma^{2}(M^{\prime})\right]}{\sigma^{2}(M)} \, .
\end{equation}
For the unrealistic case of the Markovian random walks corresponding to the choice of the sharp $k$-space filter, we expect $\kappa=0$. 
For the realistic case of the non-Markovian random walks, the exact value of $\kappa$ is dependent on the shape of the filter and very weakly on 
the background cosmology as well \citep{RL20a,RL20b,ryu-etal20}.  \citet{MR10a} proposed a simple formula of $\kappa=0.4592-0.0031R_{f}$
as a good approximation to Equation (\ref{eqn:kappa}) in the high-mass limit $M\ge 10^{14}\,h^{-1}\,M_{\odot}$. 

The obvious upside of the \mr\, mass function based on Equation (\ref{eqn:mr}) is that it is a physical model characterized by only one adjustable 
parameter, $D_{B}$, whose best-fit value was found to weakly vary with $z$ and $\Omega_{m}$. However, as explicitly admitted by \citet{MR10a}, 
the \mr\ formula for the virial mass function is valid only in the high-mass section ($M\ge 10^{14}\,h^{-1}M_{\odot}$) where the collapse dynamics is 
expected to follow well the top-hat spherical dynamics \citep{ber94}. 
Extending the \mr\ formalism to the low-mass section by incorporating the elliptical collapse dynamics and allowing the average height of the stochastic 
absorbing barrier to vary with the mass scale, i.e., $\langle\delta_{c}\rangle(M)$, \citet{CA11a} analytically derived the following 
multiplicity function 
\begin{eqnarray}
\label{eqn:ca}
f(\sigma ; D_{B},\beta) &\approx& \frac{\delta_{sc}}{\sigma} \sqrt{\frac{2}{\pi(1+D_B)}} \Bigg{\{}\exp\left[-\frac{(\delta_{sc}+\beta \sigma^2)^2}{2\sigma^2(1+D_B)}\right] \nonumber \\ 
&& -\, \frac{\kappa}{1+D_B} \Bigg{[} \biggl( 1+\frac{\beta \delta_{sc}}{1+D_B} \biggl) 
\Bigg{(}\exp\left[-\frac{\delta_{sc}^2}{2\sigma^2(1+D_B)}\right] 
-\frac{1}{2}\Gamma\bigg{[}0,\frac{ \delta_{sc}^2}{2 \sigma^2(1+D_B)}\bigg{]}\Bigg{)}  \nonumber \\ 
&& +\, \frac{\beta^2}{2(1+D_B)} \Bigg{(} \exp\left[-\frac{\delta_{sc}^2}{2\sigma^2(1+D_B)}\right]
\biggl( \sigma^2 - \frac{\delta_{sc}^2}{1+D_B}  \biggl) \nonumber \\ 
&& +\, \frac{3 \delta_{sc}^2}{4(1+D_B)} \Gamma\biggl[0,\frac{ \delta_{sc}^2}{2 \sigma^2(1+D_B)}\biggl]\Bigg{)}  \Bigg{]} \Bigg{\}} \nonumber \\
&& -\, 
\frac{\beta \kappa \delta_{sc}}{(1+D_B)^2} \biggl( 1 - \frac{\beta \delta_{sc}}{1+D_B} \biggl) \textrm{erfc}\left(\frac{\delta_{sc}}{\sigma}\sqrt{\frac{1}{2(1+D_B)}}\right)\, ,
\end{eqnarray}
where the drifting average coefficient, $\beta$, is defined as $\beta \equiv \sigma^{-1}(M)\left[\langle\delta_{c}(M)\rangle -\delta_{sc}\right]$, quantifying 
the deviation of the average of the absorbing barrier from the spherical average height of $\delta_{sc}=1.686$ in the high mass limit, i.e., 
$\lim_{M\to\infty}\langle\delta_{c}\rangle(M)=\delta_{sc}=1.686$. Regarding $\kappa$, \citet{CA11b} used a constant value of $\kappa=0.475$, 
claiming that it provides a fairly good approximation to Equation (\ref{eqn:kappa}) not only in the high-mass but also in the low-mass range 
$M\le 10^{14}\,h^{-1}\,M_{\odot}$ provided that the background is described by the $\Lambda$CDM cosmology and that the top-hat filter is used to smooth 
the density field. Throughout this paper, we adopt this approximation for $\kappa$. 

At the cost of introducing one more adjustable parameters, $\beta$, the \ca\, mass function based on Equation (\ref{eqn:ca}) succeeded in matching 
the numerical results in a much wider range of $M$ than the \mr\ virial mass function based on Equation (\ref{eqn:mr}) \citep{CA11a,CA11b}. 
Given that Equation (\ref{eqn:ca}) is a physical model derived in the generalized excursion theory, we suggest that it should be applicable to the 
splashback mass function. In Section \ref{sec:test}, we will incorporate the self-similar spherical infall dynamics into  Equation (\ref{eqn:ca}) to 
accommodate the splashback mass function and test the modified \ca\ formula against N-body simulations. 

Before proceeding to derive the splashback mass function in the generalized excursion set theory, however, we would also like to briefly review 
the multiplicity function proposed by \citet{tin-etal08}, on which the formula of \citet{die20a} is based. 
\begin{equation}
\label{eqn:tinker}
f(\sigma) = A \left[ 1+  \left(\frac{\sigma}{b} \right)^{-a} \right] \exp\left(-\frac{c}{\sigma^2} \right)\, ,
\end{equation}
where $A,\ a,\ b,$ and $p$ are four adjustable parameters. Although \citet{tin-etal08} used the same excursion set ansatz, 
i.e., Equations (\ref{eqn:massft})-(\ref{eqn:diffvol}), for the evaluation of the virial mass function with this multiplicity function, 
Equation (\ref{eqn:tinker}) is not a physical model but a mere fitting formula, unlike Equations (\ref{eqn:mr}) and (\ref{eqn:ca}). 

\subsection{Absorbing barrier for the splashback mass function}\label{sec:absorbing}

For the derivation of the splashback mass function in the generalized excursion set theory, it is first necessary to determine the spherical average 
of the stochastic absorbing barrier in the high-mass limit. For the virial mass function, the top-hat solution, $\delta_{sc}=1.686$ is used, under the 
simplified assumption that a bound halo forms through the gravitational collapse of an {\it isolated} uniform sphere into a singularity. 
For the splashback mass function, however, this top-hat solution should not be used since the concept of the splashback radii has a theoretical 
basis on the self-similar spherical infall model which is different from the top-hat spherical model \citep{FG84,ber85}.  
From here on, we let $\delta_{sp}$ denote the spherical average height of the stochastic absorbing barrier in the high mass limit for the splashback 
mass function to distinguish it from that for the virial mass function, $\delta_{sc}=1.686$. 

It should be quite reasonable and adequate to set $\delta_{sp}$ at the solution of the self-similar spherical infall model, which in fact has been already found 
by \citet{sha-etal99} to be $1.52$. Describing a bound halo as a "truncated, non-singular, isothermal sphere in virial and hydrostatic equilibrium" rather 
than an isolated singular uniform sphere, \citet{sha-etal99} analytically proved that the self-similar spherical infall solution of $\delta_{sp}=1.52$ is 
equivalent to the linear density contrast within a boundary at which the halo potential energy reaches its minimum value 
i.e., splashback boundary \citep[see Section 7.1 in][]{sha-etal99}. 

Replacing $\delta_{sc}=1.686$ by $\delta_{sp}=1.52$ in Equation (\ref{eqn:ca}) and regarding $M$ as a splashback mass rather than a virial mass 
in Equations (\ref{eqn:massft})-(\ref{eqn:diffvol}), we suggest that the most general analytic form for the splashback mass function in the excursion 
set theory should be written as 
\begin{eqnarray}
\label{eqn:ca_mod}
\frac{dN(M, z)}{d\ln M} &=&  \frac{\bar{\rho}}{M}\Bigg{\vert}\frac{d\ln\sigma^{-1}}{d\ln M}\Bigg{\vert}
\frac{\delta_{sp}}{\sigma} \sqrt{\frac{2}{\pi(1+D_B)}} \Bigg{\{}\exp\left[-\frac{(\delta_{sp}+\beta \sigma^2)^2}{2\sigma^2(1+D_B)}\right] \nonumber \\ 
&& -\, \frac{\kappa}{1+D_B} \Bigg{[} \biggl( 1+\frac{\beta \delta_{sp}}{1+D_B} \biggl) 
\Bigg{(}\exp\left[-\frac{\delta_{sp}^2}{2\sigma^2(1+D_B)}\right] 
-\frac{1}{2}\Gamma\bigg{[}0,\frac{ \delta_{sp}^2}{2 \sigma^2(1+D_B)}\bigg{]}\Bigg{)}  \nonumber \\ 
&& +\, \frac{\beta^2}{2(1+D_B)} \Bigg{(} \exp\left[-\frac{\delta_{sp}^2}{2\sigma^2(1+D_B)}\right]
\biggl( \sigma^2 - \frac{\delta_{sp}^2}{1+D_B}  \biggl) \nonumber \\ 
&& +\, \frac{3 \delta_{sp}^2}{4(1+D_B)} \Gamma\biggl[0,\frac{ \delta_{sp}^2}{2 \sigma^2(1+D_B)}\biggl]\Bigg{)}  \Bigg{]} \Bigg{\}} \nonumber \\
&& -\, 
\frac{\beta \kappa \delta_{sp}}{(1+D_B)^2} \biggl( 1 - \frac{\beta \delta_{sp}}{1+D_B} \biggl) \textrm{erfc}\left(\frac{\delta_{sp}}{\sigma}\sqrt{\frac{1}{2(1+D_B)}}\right)\, ,
\end{eqnarray}
and call Equation (\ref{eqn:ca_mod}) the modified \ca\ model characterized by two free parameters, $D_{B}$ and $\beta$. 

If the drifting average coefficient becomes negligibly small, i.e., $\beta=0$, for the case of the halos identified by the splashback raii,  
Equation (\ref{eqn:ca_mod}) can be reduced to the modified \mr\, mass function characterized by only one free parameter, $D_{B}$: 
\begin{eqnarray}
\label{eqn:mr_mod}
\frac{dN(M, z)}{d\ln M} &=&  \frac{\bar{\rho}}{M}\Bigg{\vert}\frac{d\ln\sigma^{-1}}{d\ln M}\Bigg{\vert}\frac{\delta_{sp}}{\sigma}
\sqrt{\frac{2}{\pi(1+D_{B})}}\Bigg{\{}\left[1-\frac{\kappa}{(1+D_{B})}\right]
\exp\left[-\frac{\delta_{sp}^2}{2\sigma^2(1+D_{B})}\right]  \nonumber \\
&&+\frac{\kappa}{2(1+D_B)}\Gamma\left[0,\frac{\delta_{sp}^2}{2 \sigma^2(1+D_{B})}\right]\Bigg{\}}\, ,
\end{eqnarray}
where $\kappa$ is given as Equation (\ref{eqn:kappa}) but with splashback masses $M$ and $M^{\prime}$. 
 
If the halo identification by the splashback boundary becomes unambiguous, Equation (\ref{eqn:ca_mod}) could be even further reduced to 
the following simple parameter-free formula: 
\begin{equation}
\label{eqn:para_free}
\frac{dN(M, z)}{d\ln M} =  \frac{\bar{\rho}}{M}\Bigg{\vert}\frac{d\ln\sigma^{-1}}{d\ln M}\Bigg{\vert}\frac{\delta_{sp}}{\sigma}
\sqrt{\frac{2}{\pi}}\Bigg{\{}\left(1-\kappa\right)
\exp\left(-\frac{\delta_{sp}^2}{2\sigma^2}\right)+\frac{\kappa}{2}\Gamma\left[0,\frac{\delta_{sp}^2}{2 \sigma^2}\right]\Bigg{\}}\, .
\end{equation}
In Section \ref{sec:numerical}, we will first compare Equation (\ref{eqn:ca_mod}) with the numerically obtained splashback mass functions to determine 
the values of $D_{B}$ and $\beta$. Then, we will investigate under what conditions the splashback mass function can be described by the single 
parameter model, Equation (\ref{eqn:mr_mod}) or by the parameter free model, Equation (\ref{eqn:para_free}). 

\section{Numerical Testing of the Analytical Models}\label{sec:numerical}

\citet{sparta17} developed an elaborate code, called SPARTA, to efficiently locate the splashback boundaries of DM halos. This code basically 
follows the orbits of all DM particles in and around a given halo identified by the Rockstar halo-finder algorithm \citep{rockstar} and determines 
the distances to the apocenters of their first orbits, $r_{\rm sp}$.  To derive the splashback radius of each halo, $R_{\rm sp}$,  from the distribution of 
$r_{\rm sp}$, \citet{sparta17} considered  several different definitions. For instance, $R_{\rm sp}^{\rm min}$ is the splashback radius of a halo defined as 
the average of the $r_{\rm sp}$ values, while $R_{\rm sp}^{Q\%}$ is defined as the Q-th percentile of the distribution of $r_{\rm sp}$.  
For our analysis, we adopt mainly the former definition, $R_{\rm sp}^{\rm min}$, and refer to the mass enclosed by $R_{\rm sp}^{\rm min}$ as the 
splashback mass of a halo, unless otherwise stated. 

\citet{die-etal17} applied this code to the Erebos suite of 14 DM only N-body simulations for three different (Planck, WMAP7 and self-similar) 
cosmologies and produced the halo catalogs \footnote{They are available at the website of http://erebos.astro.umd.edu/erebos/ .} 
that provided additional information on the splashback radius and mass of each halo \citep[see also][]{die20b}. 
From the Erebos suite, we utilize two simulations of $1024^{3}$ DM particles run by \citet{DK14} and \citet{DK15} with the GADGET2 code
\citep{gadget2} on a periodic box of linear size $500\,h^{-1}$Mpc for the Planck and WMAP7 cosmologies, respectively.  
Table \ref{tab:sim} lists the mass of individual DM particle ($m_{p}$) and value of $\Omega_{m}$ adopted for each simulation in the first three columns.  

From the halo catalog of each simulation at $z=0$, we select only those DM halos whose splashback masses, $M$, are equal to or larger than 
$5\times 10^{12}\,h^{-1}\,M_{\odot}$ (i.e., galaxy group and cluster scales), corresponding to the particle number cut of $500$ and $575$ for the Planck and 
WMAP7 cosmologies, respectively. 
Splitting the range of $\ln M$ into short intervals of length $d\ln M$, we determine the number densities of the 
selected DM halos, $dN/d\ln M$, at each bin.  We first fit the modified \ca\ formula to this numerically obtained $dN/d\ln M$ by adjusting 
the values of $D_{B}$ and $\beta$ via the $\chi^{2}$-minimization:  
\begin{equation}
\label{eqn:chi2}
\chi^{2}(D_{B},\beta) = \sum_{i}\frac{\left[n^{\rm num}(m_{i})-n^{\rm ana}(m_{i};D_{B},\beta)\right]^{2}}{\sigma^{2}_{n}(m_{i})}\, ,
\end{equation}
where $n^{\rm num}(m_{i})$ and $n^{\rm ana}(m_{i};D_{B},\beta)$ denote the numerical and analytical values of $dN/d\ln M$, respectively, 
at the $i$th logarithmic mass bin, $m_{i}\equiv \ln M_{i}$. Here, the error $\sigma_{n}$ at each bin is obtained through the following Jackknife analysis.  
Splitting the halo sample into eight subsamples according to the halo positions, we evaluate $dN/d\ln M$ separately for each subsample. 
Then, the Jackknife error of the splashback mass function at each logarithmic mass bin, $\sigma(m_{i})$, is obtained as one standard deviation 
scatter of $dN/d\ln M$ among the eight subsamples. 

\begin{figure}
\begin{center}
\includegraphics[scale=0.7]{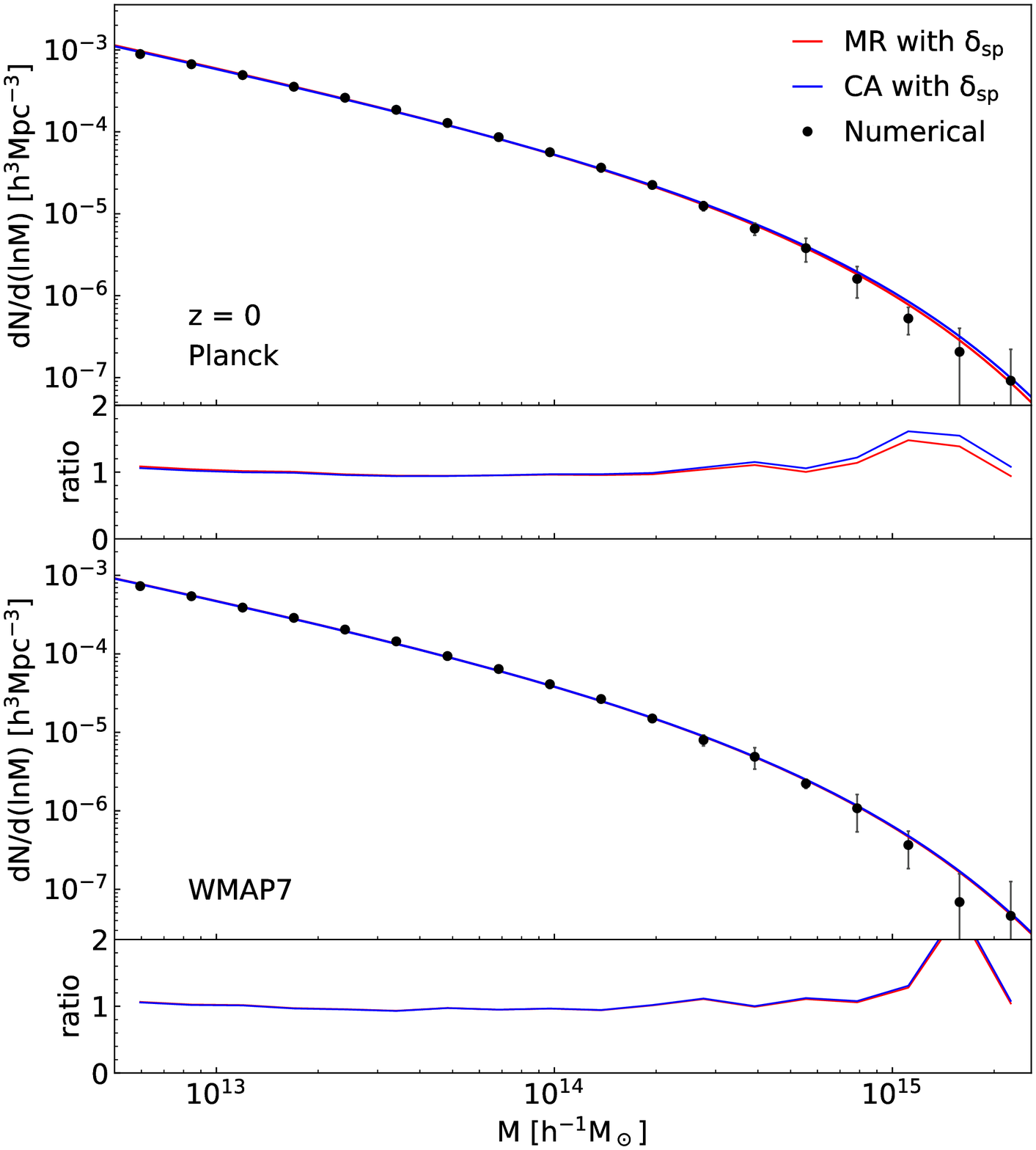}
\caption{Number densities of the DM halos as a function of their splashback masses (black filled circles) with the Jackknife errors 
from the Erebos simulations \citep{die-etal17} and the modified \ca\, formula given in Equation (\ref{eqn:ca_mod}) for the splashback 
mass function (red solid lines) at $z=0$ for two different $\Lambda$CDM cosmologies.}
\label{fig:ca}
\end{center}
\end{figure}

Figure \ref{fig:ca} plots the results (black filled circles) with the Jackknife errors as well as the modified \ca\, mass function (blue solid lines), 
Equation (\ref{eqn:ca_mod}), with the best-fit values of $D_{B}$ and $\beta$ at $z=0$ for the Planck and WMAP7 cosmologies 
(in the first and third panels from the top, respectively). The ratio of the analytical to the numerical splashback mass functions, 
$n^{\rm ana}/n^{\rm num}$, are also shown for the two cosmologies (in the second and fourth panels from the top, respectively). 
As can be seen, the modified \ca\ mass function indeed agrees very well with the numerical results, exhibiting, $1\le n^{\rm ana}/n^{\rm num}\le 1.1$ 
in the mass range of $0.5\le M/(10^{13}\,h^{-1}\,M_{\odot})\le 50$ for both of the cosmologies. 
In the higher mass range of $M/(10^{13}\,h^{-1}\,M_{\odot})> 50$ where the numerical results suffer from larger errors, 
we note a substantial discrepancy, $n^{\rm ana}/n^{\rm num}\ge 1.5$. 

\begin{figure}
\begin{center}
\includegraphics[scale=0.7]{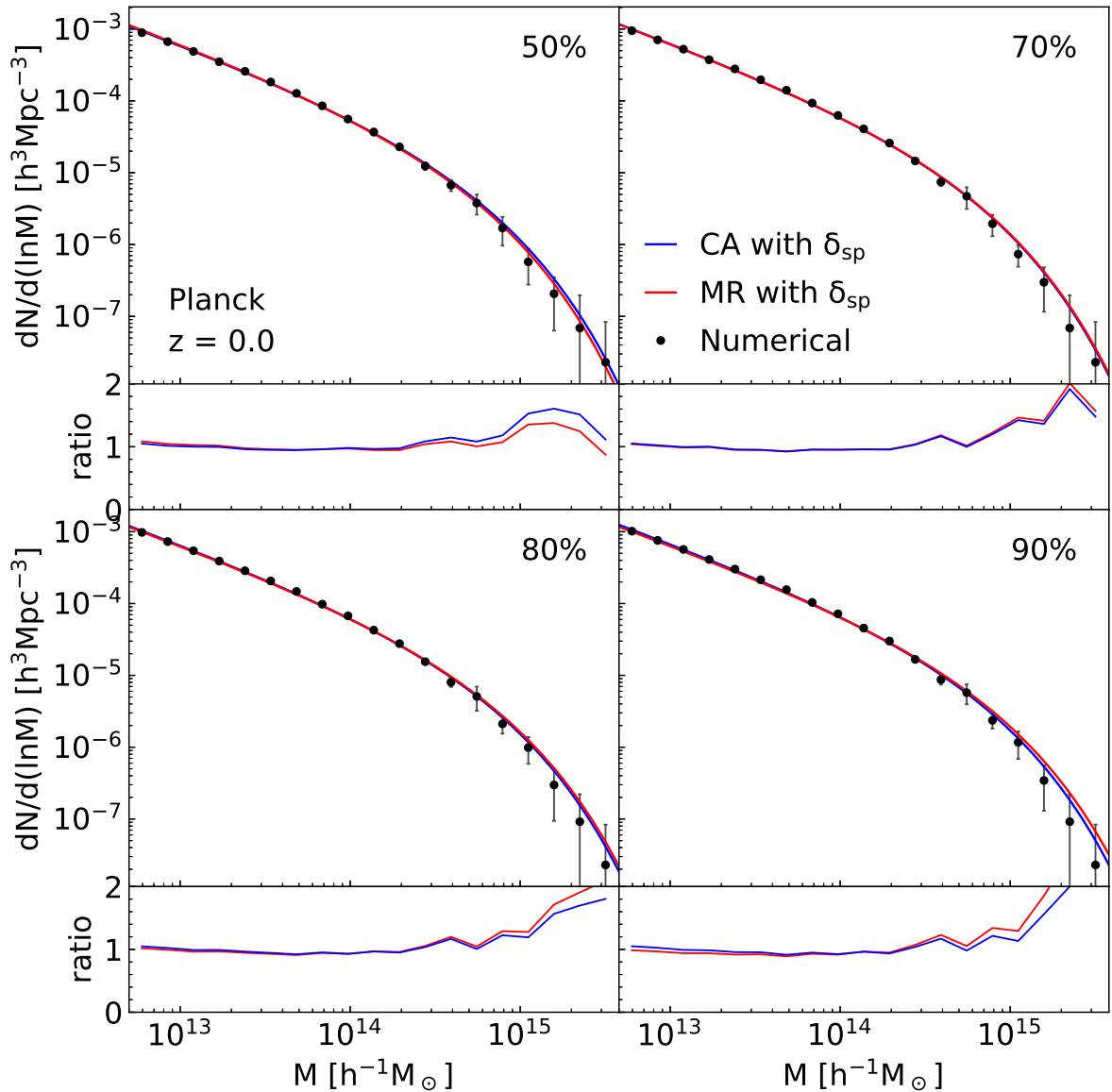}
\caption{Same as Figure \ref{fig:ca} but for the case that the splashback radius is defined as 
$R_{\rm sp}^{50\%}$ (top-left), $R_{\rm sp}^{70\%}$ (top-right), $R_{\rm sp}^{80\%}$ (bottom-left) 
$R_{\rm sp}^{90\%}$ (bottom-right) for the Planck cosmology.}
\label{fig:per_planck}
\end{center}
\end{figure}

\begin{figure}
\begin{center}
\includegraphics[scale=0.7]{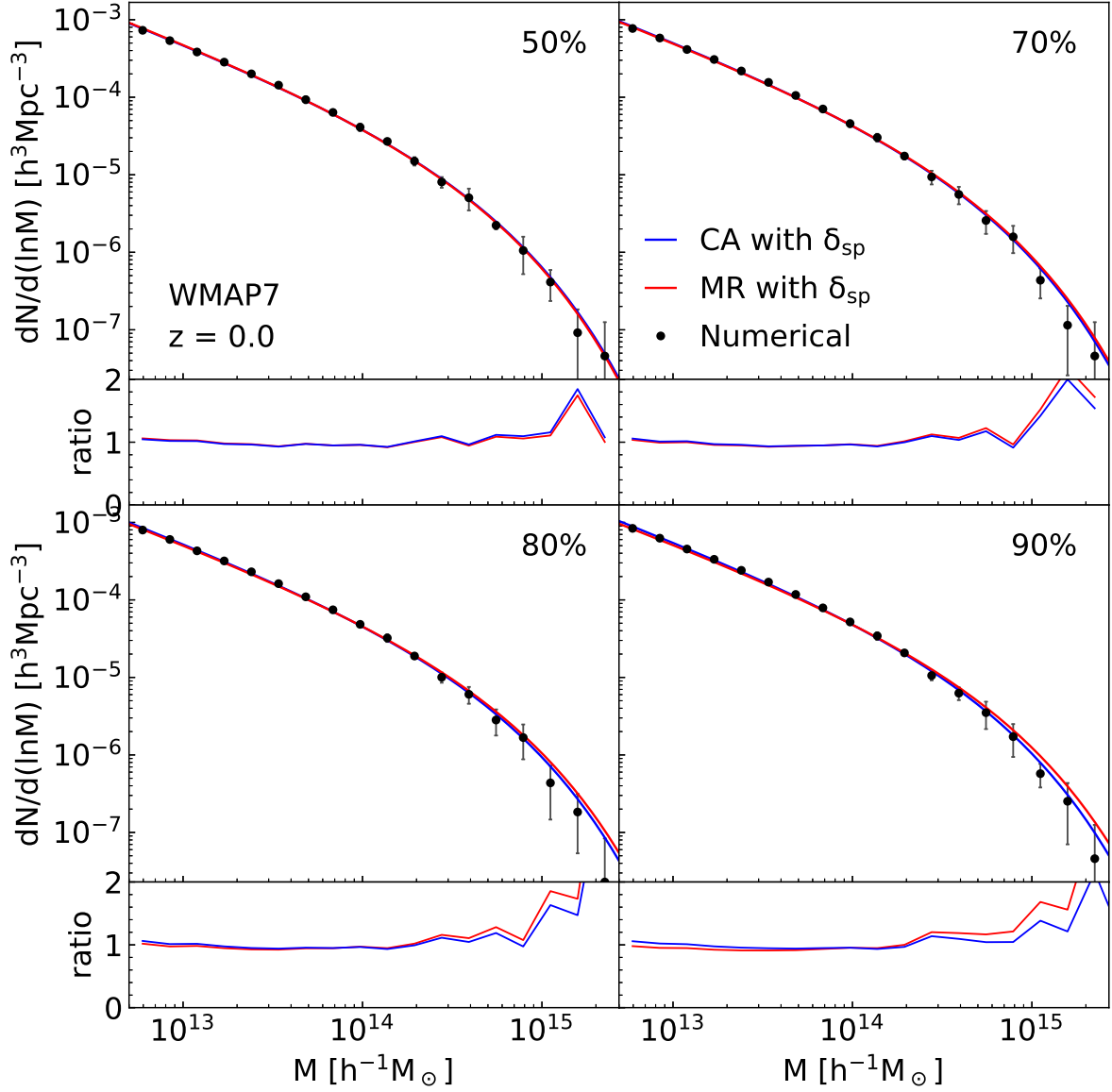}
\caption{Same as Figure \ref{fig:per_planck} but for the WMAP7 cosmology.}
\label{fig:per_wmap7}
\end{center}
\end{figure}

We investigate if this analytical splashback mass function based on the generalized excursion set theory is valid even for the case that different definitions 
of $R_{\rm sp}$ are used to determine the splashback masses. Figures \ref{fig:per_planck} and \ref{fig:per_wmap7} plot the same as Figure \ref{fig:ca} 
but for four different cases that the splashback radii are defined by $R_{\rm sp}^{50\%}$ (top-left panel), $R_{\rm sp}^{70\%}$ (top-right panel), 
$R_{\rm sp}^{80\%}$ (bottom-left panel) and $R_{\rm sp}^{90\%}$ (bottom-right panel) for the Planck and WMAP7 cosmologies, respectively. 
As can be seen,  the validity of the \ca\, formula in the mass range of $0.5\le M/(10^{13}\,h^{-1}\,M_{\odot})\le 50$ is fairly 
robust against the variation of the definition of $R_{\rm sp}$ for both of the cosmologies. 
It is, however, worth mentioning that for these plots, we newly determine the best-fit values of $D_{B}$ and $\beta$ for each case via Equation (\ref{eqn:chi2}). 
That is, the two parameters, $D_{B}$ and $\beta$, are dependent on the percentiles, $Q$. For the case of $R_{\rm sp}^{50\%}$ the best-fit values are found 
to be almost identical to those for the case of $R^{\rm min}_{\rm sp}$, while the other three cases yield substantially different best-fit parameters.  
We do not pursue here an analytical modeling of the percentile dependence of the two parameters and consistently use $R_{\rm sp}^{\rm min}$ as a 
definition of the splashback radius. 

We investigate if and how the splashback mass functions differ in the best-fit values of $D_{B}$ and $\beta$ from the virial mass functions, for which  
we numerically determine $dN/d\ln M$ as a function of their virial masses and analytically evaluate the original \ca\ formula with $\delta_{sc}=1.686$. 
Figure \ref{fig:cont} plots the $68\%,\ 95\%$ and $99\%$ contours of $\chi^{2}$ (red, blue and green lines, respectively) in the $\beta$-$D_{B}$ configuration 
space at six different redshifts in the range of $0\le z\le 2$ for the two cosmologies.  As can be seen, the values of $\beta$ and $D_{B}$ for the splashback mass 
functions (solid lines) are progressively lower than those for the virial mass functions (dotted lines), which implies the following:  For the case that the splashback 
rather than the virial radius is used as the halo boundary, the formation process of a DM halo follows much better the spherical collapse 
dynamics and thus its identification becomes much less ambiguous. 

Noting that the best-fit values of the drifting average coefficient, $\beta$, shown in Figure \ref{fig:cont} are quite close to zero at all of the redshifts considered, 
we set it at zero and suggest that the modified \mr\, formula, Equation (\ref{eqn:mr_mod}), with only one parameter  $D_{B}$ should be a 
more efficient model for the splashback mass function. Figure \ref{fig:ca} also plots the modified \mr\, formula with the best-fit value of $D_{B}$ 
(red solid lines) and its ratio to the numerical results. The comparisons of the modified \mr\ and \ca\ formulae with the numerical results at three 
higher redshifts in the range of $0\le z\le 2$ can be witnessed in Figure \ref{fig:mr}. 
The results shown in Figures \ref{fig:ca} and \ref{fig:mr} demonstrate that even though it has only one free parameter, the modified \mr\, formula can describe the 
numerically obtained splashback mass function as validly as the modified \ca\ formula in the redshift range of $0\le z\le 2$, which justifies our adoption of 
$\beta=0$. 
We also test and confirm the robustness of the \mr\, formula against the definitions of $R_{\rm sp}$, the results of which are shown (red solid lines) in 
Figures \ref{fig:per_planck}-\ref{fig:per_wmap7}. 

It is worth recalling the numerical finding of \citet{gar-etal21} that when the halo boundaries were defined as the radii that optimize the halo fits, the 
Press-Schechter like formula with a modified spherical barrier height of $\delta_{sc}=1.45$ lower than the top-hat solution of $\delta_{sc}=1.68$ matched 
quite well the numerically obtained halo mass function. Given that the halo boundaries defined by \citet{gar-etal21} are slightly larger than the splashback radii, 
the success of the modified \mr\ formula with $\delta_{sp}=1.52$ shown in Figures \ref{fig:ca}-\ref{fig:mr} is in line with their findings, providing a physical 
explanation to them.  
 
Figure \ref{fig:Db_z} plots the resulting best-fit values of $D_{B}$ (filled circles) as a function of $z$ for both of the cosmologies. Here, the error 
in the best-fit value of $D_{B}$ is found as the inverse of the square root of the Fisher information. 
As can be seen, the best-fit value of $D_{B}$ appears to decrease almost linearly with the increment of $z$ toward zero, 
which motivates us to model $D_{B}(z)$ by the following linear function of $z$, 
\begin{equation} 
\label{eqn:Db_z}
D_{B}(z) = A_{D}(z_{c}-z)\, ,
\end{equation} 
characterized by two parameters, the slope, $A_{D}$, and threshold redshift, $z_{c}$, satisfying $A_{D}\equiv \vert dD_{B}(z)/dz\vert$ and 
$D_{B}(z_{c})=0$, respectively.  Employing the $\chi^{2}$-minimization method, we statistically compare Equation (\ref{eqn:Db_z}) with the 
numerically obtained $D_{B}(z)$ to find the best-fit values of $A_{D}$ and $z_{c}$ as well as their errors for the two cosmologies, which are listed 
in the fourth and fifth columns of Table \ref{tab:sim}. Note that a significant difference in $z_{c}$ but no difference in $A_{D}$ are found between the two 
$\Lambda$CDM cosmologies, which hints that $z_{c}$ may sensitively depend on $\Omega_{m}$.  

The result shown in Figure \ref{fig:Db_z} also indicates that the absorbing barrier for the splashback mass function becomes less and less stochastic as 
$z$ increases, until it eventually turns to being deterministic at $z=z_{c}$. Figure \ref{fig:para_free} compares the numerical results (black filled circles) 
with the parameter free formula given in Equation (\ref{eqn:para_free}) at $z=z_{c}$ for the two cosmologies. As can be seen, good agreements are 
found between them ($1\le n^{\rm ana}/n^{\rm num}< 1.1$) in the mass range of $0.5\le M/(10^{13}\,h^{-1}\,M_{\odot})\le 2$ for both of the cosmology, 
even though no empirical adjustment is made at all. Yet, in the higher mass section, the large errors make it difficult to make a proper comparison. 

We have so far considered a rather restricted mass-range of $0.5\le M/(10^{13}\,h^{-1}\,M_{\odot})\le 10^{2}$, mainly because of the limited mass-resolution of the 
simulation box of $L_{\rm box}=500\,h^{-1}$Mpc and partially because the splashback mass functions are most useful as a cosmological diagnostics on the galaxy 
group and cluster scales. To examine if the validity of the modified \mr\ formula may also be extended to a wider mass range, we repeat the same analysis but with 
the data obtained from the simulation boxes of two different linear sizes, $L_{\rm box}=125$ and $250\,h^{-1}$Mpc for both of the cosmologies. 
Two additional datasets from the larger simulation boxes with $L_{\rm box}=1000$ and $2000\,h^{-1}$Mpc are also available for the WMAP7 cosmologies. 
For a detailed information on these simulations with various $L_{\rm box}$, see \citet{sparta17}. 

\begin{figure}
\begin{center}
\includegraphics[scale=0.7]{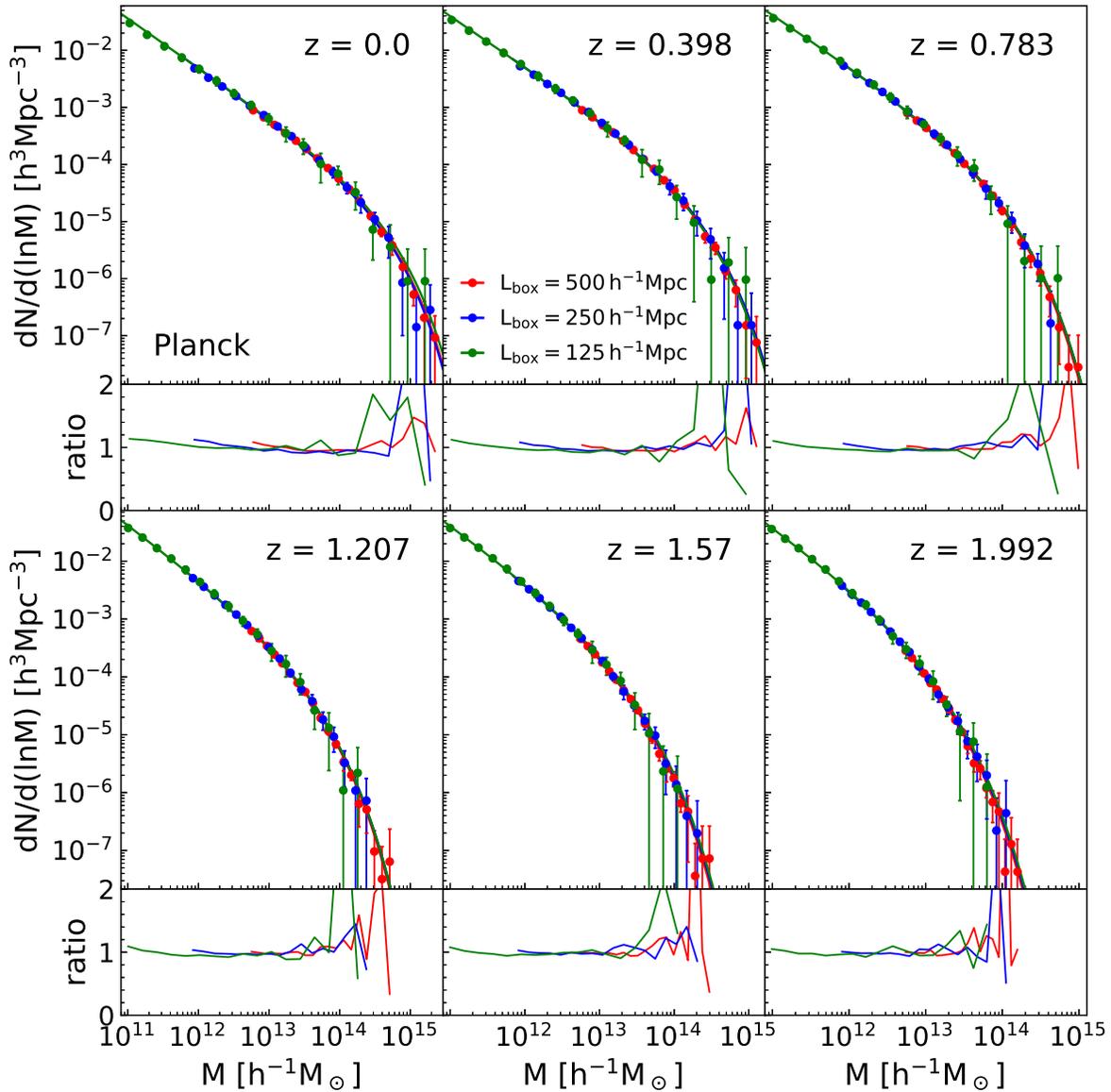}
\caption{Numerically obtained splashback mass functions (filled circles) from different simulation boxes 
compared with the modified \mr\, formula (solid lines) at six different redshifts for the Planck cosmologies. 
The best-fit value of $D_{B}$ for the analytic formula is set at the same value obtained for the case of 
$L_{\rm box}=500\,h^{-1}\,Mpc$.}
\label{fig:small_m_planck}
\end{center}
\end{figure}

\begin{figure}
\begin{center}
\includegraphics[scale=0.7]{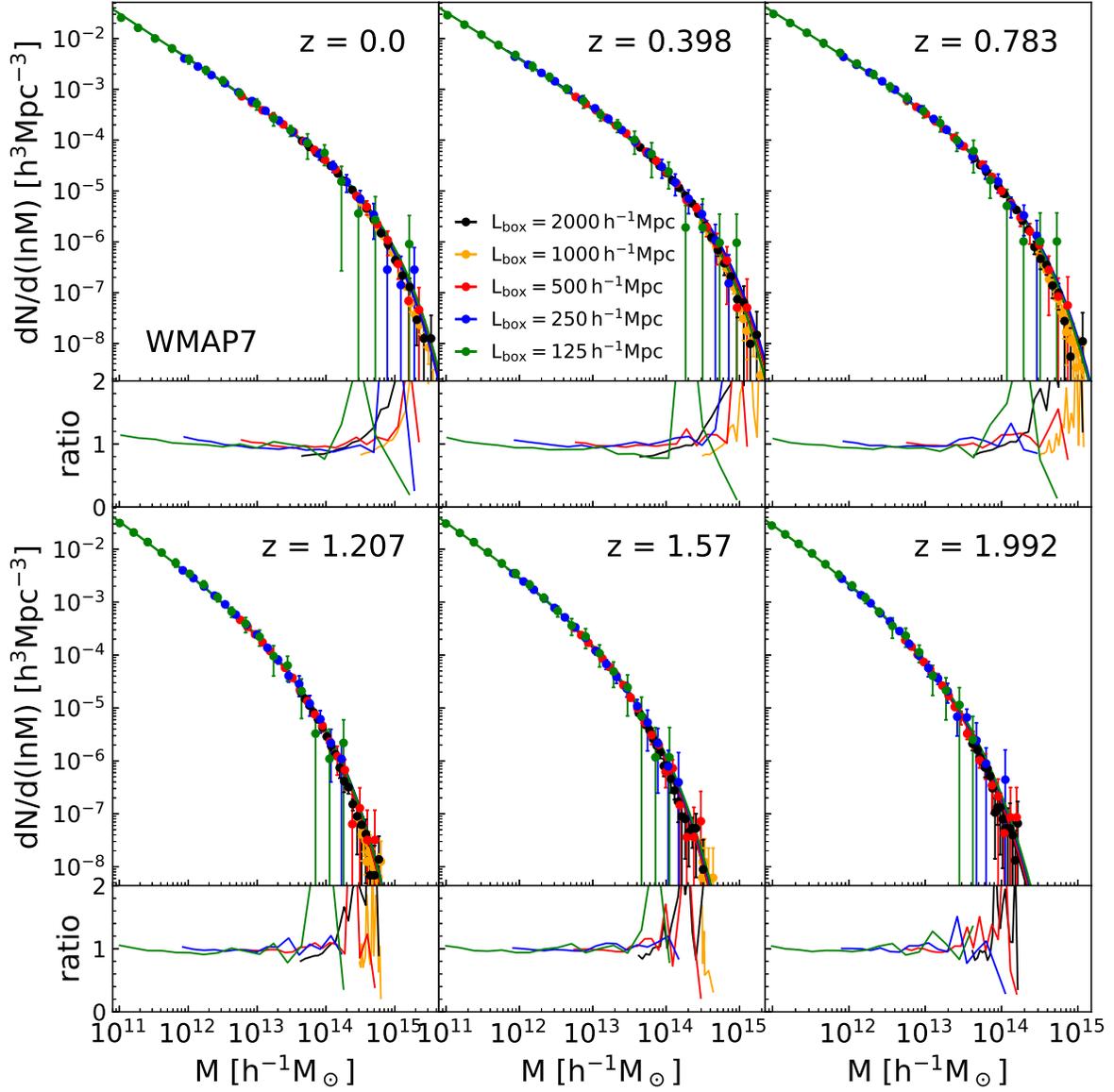}
\caption{Same as Figure \ref{fig:small_m_planck} but for the WMAP cosmology and additionally 
including the results from two larger simulation boxes.}
\label{fig:small_m_wmap}
\end{center}
\end{figure}

Figure \ref{fig:small_m_planck} and \ref{fig:small_m_wmap} compare the modified \mr\, formula with the numerical results at six different redshifts from the three 
and five different simulation boxes for the Planck and WMAP7 cosmologies, respectively.  
For the analytic formula, we do not newly determine the best-fit value of $D_{B}$ for each case of $L_{\rm box}$ but use the same best-fit value 
obtained for the case of $L_{\rm box}=500\,h^{-1}$Mpc. 
As can be seen, the modified \mr\, formula with the same best-fit value of $D_{B}$ still agrees quite well with the numerical results in the low-mass 
section of $0.1\le M/(10^{12}\,h^{-1}\,M_{\odot})\le 1$ from the smaller simulation boxes for both of the cosmologies, 
exhibiting $0.9< n^{\rm ana}/n^{\rm num} < 1.2$, even though no new adjustment of $D_{B}$ is made. 
That is, the validity of the modified \mr\ formula for the splashback mass function is not limited to the high-mass section, unlike the case of 
the virial mass function for which the \mr\ formula breaks down even in the intermediate mass section of $10^{13}\le (M/h^{-1}\,M_{\odot}) < 10^{14}$. 
Meanwhile, for the WMAP7 cosmology, in the higher mass section of  $1\le M/(10^{15}\,h^{-1}\,M_{\odot})\le 5$ where the numerical results 
carry relatively larger errors, the agreements between the modified \mr\, formula with the fixed value of $D_{B}$ and the numerical 
results from the larger simulation boxes turn out to be not so good as in the lower-mass section. 

\begin{figure}
\begin{center}
\includegraphics[scale=0.7]{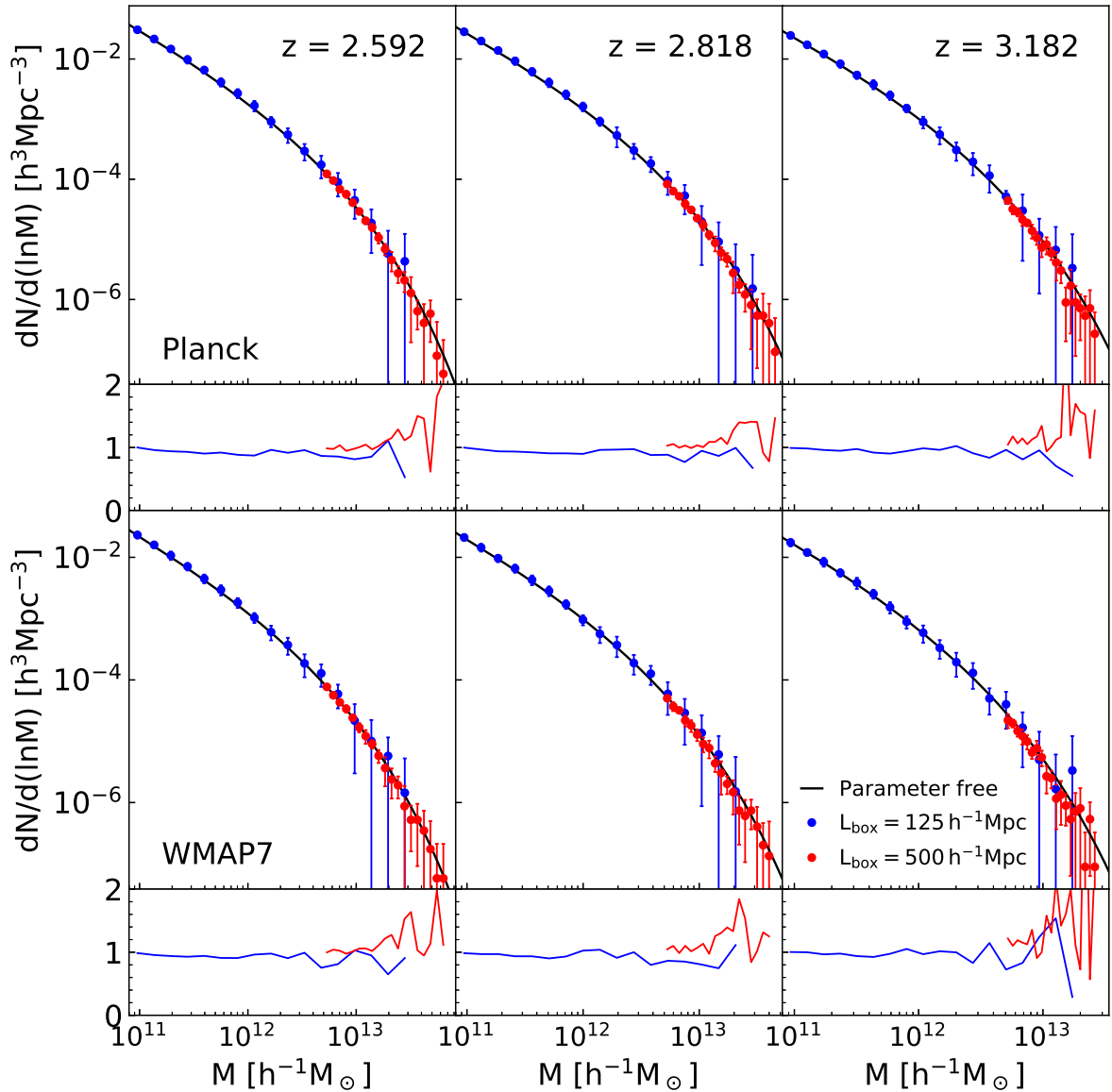}
\caption{Numerically obtained splashback mass functions from two different simulation boxes (blue and 
red filled circles) at three higher redshifts $z>z_{c}$, compared with the parameter free formula.}
\label{fig:para_free_z}
\end{center}
\end{figure}

Using the data from this smaller simulation box, we also examine whether or not at $z\ge z_{c}$ the parameter free formula, Equation (\ref{eqn:para_free}), is 
capable of matching the numerical results in a wider mass range. Figure \ref{fig:para_free_z} plots the parameter free formula and the numerical results from the simulation boxes of $L_{\rm box}=125$ and $500\,h^{-1}$Mpc at three redshifts higher than $z_{c}$ in the range of $2.5\le z\le 3.2$. 
As can be seen, in the mass range of $0.1\le M/(10^{12}\,h^{-1}\,M_{\odot})<10$  the parameter free analytic formula agrees quite well with the numerical results, 
having $1<n^{\rm ana}/n^{\rm num}<1.15$ at all of the three redshifts. This result is consistent with the aforementioned claim that the collapse barrier for the halos 
identified by the splashback radii becomes deterministic at $z\ge z_{c}$.  

Now that the modified \mr\, formula for the splashback mass function has been found to work very well in the wide ranges of $M$ and $z$, 
we would like to statistically decide which model for the splashback mass function is most preferred by the numerical data among the 
modified \ca, \mr, \st, and \tn\,  formulae.  Here, the modified \st\,  mass function is obtained by replacing $\delta_{sc}$ in Equation 
(\ref{eqn:st}) by $\delta_{sp}$ as done for the modified \ca\, and \mr\, formulae and newly determining the best-fit values of its three parameters 
for the splashback mass function. As for the modified \tn\, mass function in which no term contains $\delta_{sc}$, we obtain it by newly determining 
the best-fit values of the four parameters by fitting Equations (\ref{eqn:massft}), (\ref{eqn:diffvol}) and (\ref{eqn:tinker}) to the numerically obtained 
splashback mass function.

To select the model most preferred by the numerical results, we calculate the Bayesian information criterion (BIC) of each model, defined as 
\citep{bic} 
\begin{equation} 
\label{eqn:bic}
{\rm BIC} = N_{\rm pa}\ln N_{\rm pt} + \chi^{2}_{\rm min}\, ,
\end{equation}
where $N_{\rm pa}$ and  $N_{\rm pt}$  denote the numbers of the free parameters of each model and the number of the experimental data points, 
respectively, and $\chi^{2}_{\rm min}$ denotes the minimum value of $\chi^{2}$ for each model. The model with lower value of ${\rm BIC}$ is more 
preferred by the numerical results. 

Figure \ref{fig:bic} plots the BIC values of the modified \ca, \mr, \st, \tn\, (blue, red, green, violet sold lines, respectively) formulae as a function of $z$ 
for the two $\Lambda$CDM cosmologies. As can be seen,  the modified \mr\, formula characterized by only one single parameter, $D_{B}$, is found to 
have the smallest BIC value in the redshift range of $0.3<z\le 3$, which proves that the modified \mr\ formula is indeed the most preferred analytic model for 
the splashback mass function among the four in this redshift range. 

Yet, it should be noted that the modified \mr\ and \ca\ formulae exhibit sharp increases with the decrement of $z$ in the lower redshift range of $z\le 0.3$, 
where the modified \st\ formula yields the smallest BIC values. To understand the reason for this abrupt increases of the BIC values at $z\le 0.3$, 
we examine the $\chi^{2}$ values and find that both of the \ca\ and \mr\ formulae yield relatively high values of $\chi^{2}$ in the low mass section 
$M\le 10^{13}\,h^{-1}\,M_{\odot}$ at $z\le 0.3$, which in turn leads to produce larger $BIC$ values at these low redshifts. 
Given that the more nonlinearly evolved tidal fields at lower redshifts \citep{web96,LS10} are expected to more severely disturb the lower-mass DM halos 
\citep{ber94}, we suspect that the single diffusion coefficient $D_{B}$ introduced by the generalized excursion set theory \citep{MR10a,MR10b} is no longer 
capable of accounting for the stochasticity involved in the halo identifications due to the severe tidal disturbance at $z\le 0.3$, 
even when the splashback radii are used as the halo boundaries. 

We also want to examine which one is preferred by the numerical results between the modified \mr\  with one parameter and the universal formula with 
four parameters proposed by \citet{die20a}. 
Although the same functional form of the \tn\ multiplicity function, i.e., Equation (\ref{eqn:tinker}), was adopted by \citet{die20a} to construct 
his universal formula, the best-fit parameters found by fitting the modified \tn\ formula to the numerically obtained splashback mass function 
$dN/d\ln M$ at a fixed redshift turn out to be different from those found by fitting the universal formula to the numerically obtained $f(\nu)$ with 
$\nu\equiv \delta_{sc}/\sigma(M,z)$ over a broad range of $M$ and $z$.
For this examination, the BIC method is not applicable since it is impossible to specify the number of data points, $N_{\rm pt}$, at a fixed 
redshift for the case of the universal formula of \citet{die20a}.  Instead, we calculate the alternative Akaike Information Criterion (AIC) defined as \citep{aic} 
\begin{equation}
\label{eqn:aic}
{\rm AIC} \equiv 2N_{pa} + \chi^{2}\, ,
\end{equation}
and known to complement the BIC method in case that the fitting procedure is performed over a different range of an independent variable 
between the two target models.  

\begin{figure}
\begin{center}
\includegraphics[scale=0.7]{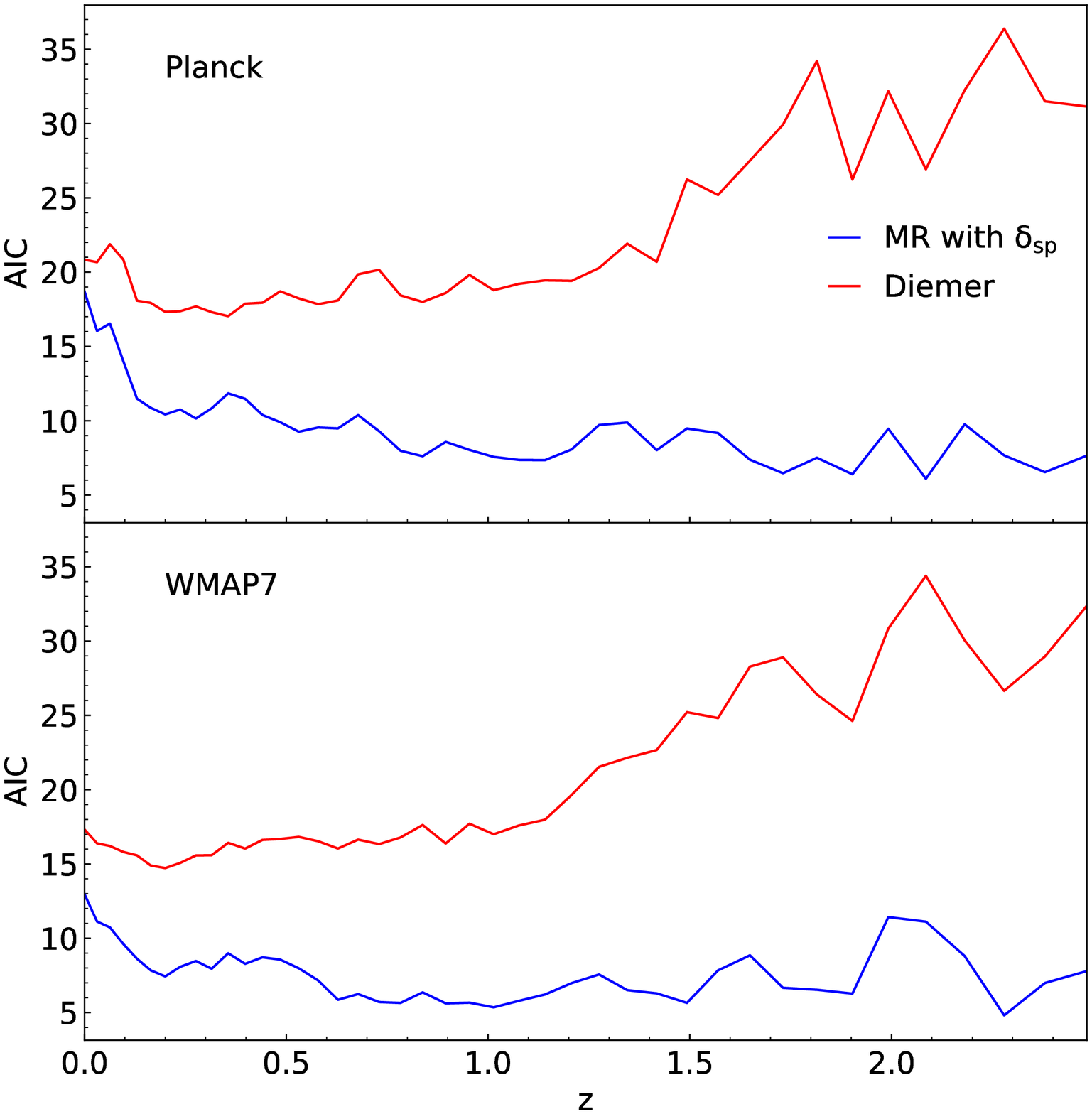}
\caption{Akaike information criterion values of $2$ different analytic models of the splashback mass function 
versus redshifts for the two different cosmologies. }
\label{fig:aic}
\end{center}
\end{figure}

Figure \ref{fig:aic} plots the AIC values of the modified \mr\ (red solid lines) and the universal formula of \citet{die20a} (blue solid lines) in the redshift range 
of $0\le z\le 3$ for the two $\Lambda$CDM cosmologies.  For the latter, we adopt the same best-fit values of the four parameters as given in \citet{die20a}. 
As can be seen, the modified \mr\ formula yields a much lower value of AIC than the universal formula of \citet{die20a} in this redshift range, 
indicating that the former is more preferred by the numerical results at least in this redshift range for the two $\Lambda$CDM cosmologies. 

 \section{Discussion and Conclusion}\label{sec:con}
 
\begin{figure}
\begin{center}
\includegraphics[scale=0.7]{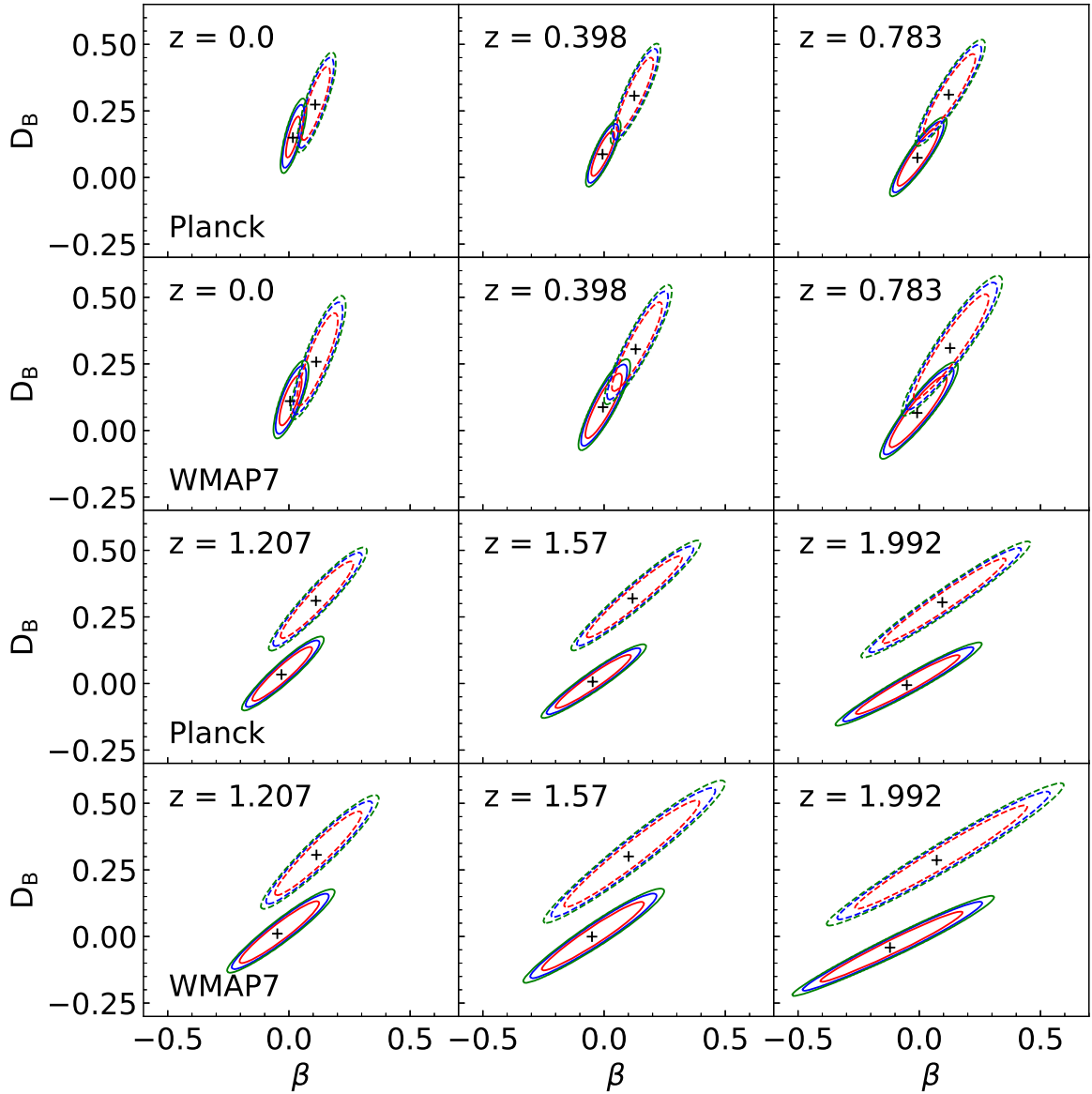}
\caption{$68\%,\ 95\%$ and $99\%$ contours of $\chi^{2}(\beta,\ D_{B})$ for the cases of the splashback (solid lines) 
and virial (dotted lines) mass functions at six different redshifts for the two cosmologies.}
\label{fig:cont}
\end{center}
\end{figure}

\begin{figure}
\begin{center}
\includegraphics[scale=0.7]{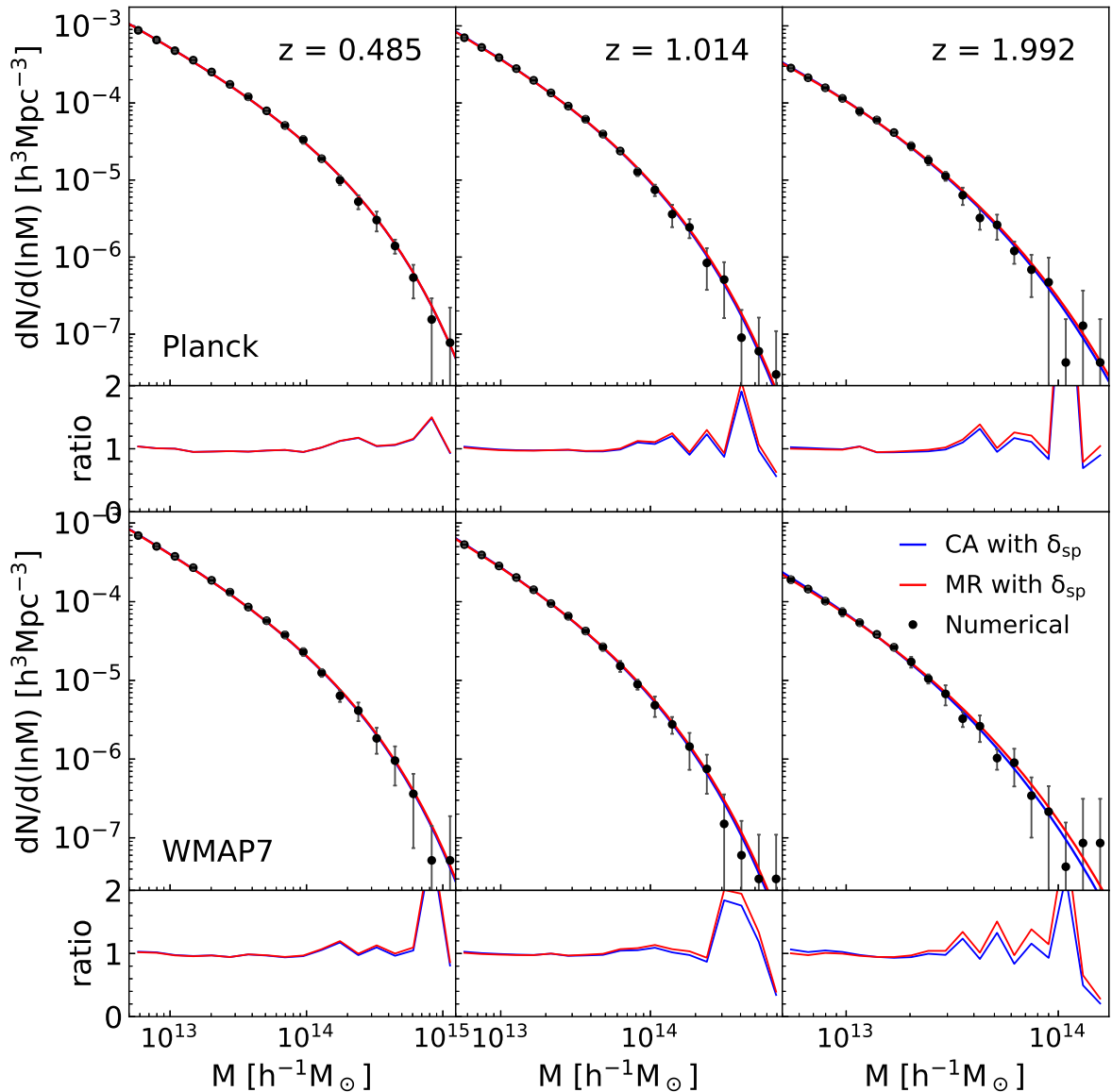}
\caption{Numerically obtained splashback mass functions (black filled circles) with the Jackknife errors 
as well as the modified \ca\ (blue solid lines) and \mr\ (red solid lines) formulae at three different redshifts 
for the two cosmologies.}
\label{fig:mr}
\end{center}
\end{figure}

We have constructed an efficient physical model for the splashback mass function in the framework of the generalized excursion set theory.  
Noting that for the cases of the halos identified by their splashback radii the absorbing barrier of the excursion set formalism 
has a lower degree of stochasticity without drifting on average from the spherical height (Figure \ref{fig:cont}), 
we have modified the \mr\, formula \citep{MR10a,MR10b} by replacing the top-hat solution of the spherical height, $\delta_{sc}=1.686$, 
by the self-similar spherical infall solution, $\delta_{sp}=1.52$ \citep{sha-etal99}. 
The modified \mr\, formula, Equation (\ref{eqn:mr_mod}), has been found to be in good accord with the numerically obtained splashback 
mass function from the Erbos simulations of linear size $500\,h^{-1}$Mpc at $0\le z\le 3$ on the galaxy group and cluster scales 
(e.g., $0.5\le M/[10^{13}\,h^{-1}\,M_{\odot}]<10^{2}$ at $z=0$) for both of the Planck and WMAP7 cosmologies (Figure \ref{fig:mr}).  

\begin{figure}
\begin{center}
\includegraphics[scale=0.7]{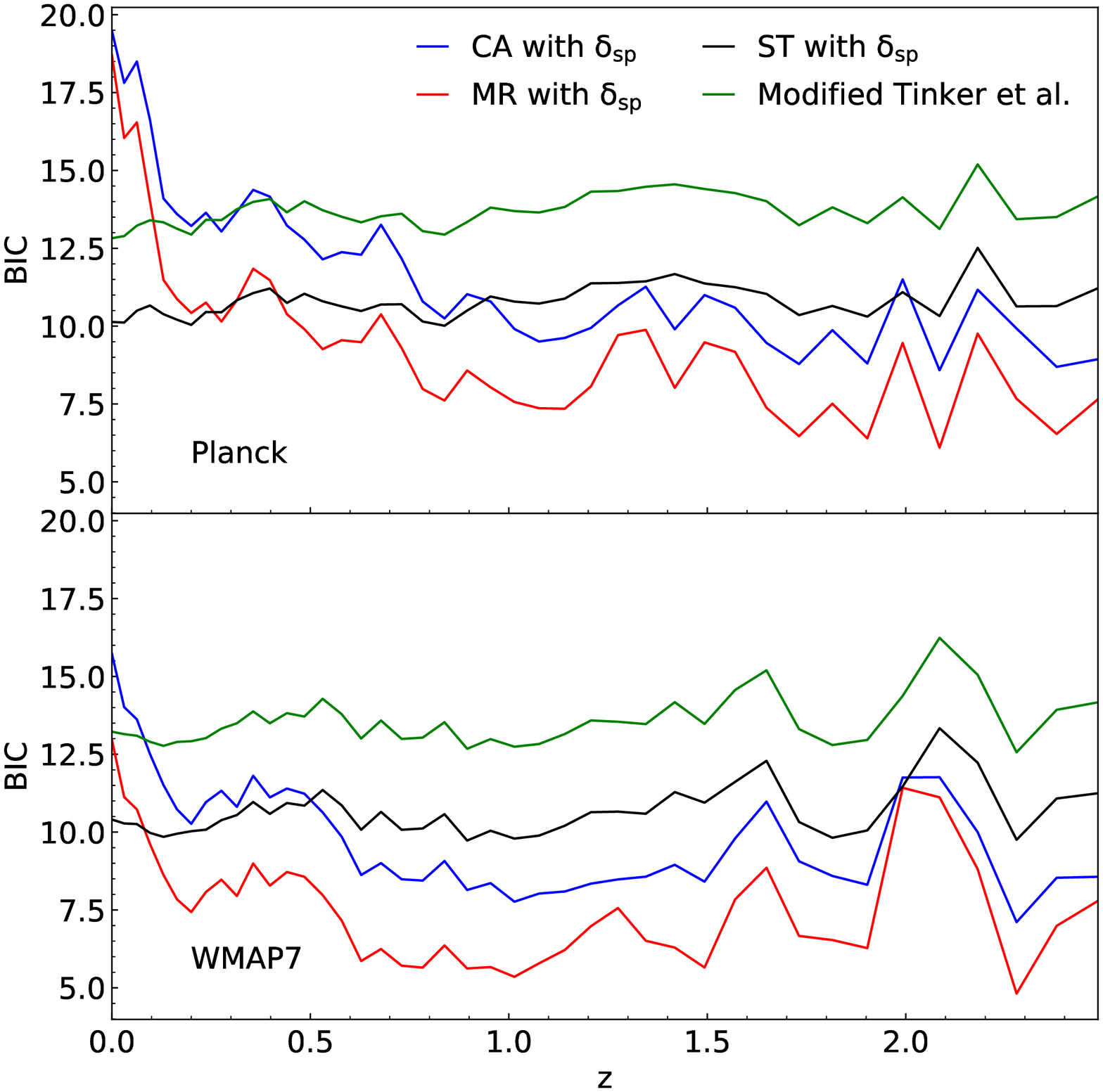}
\caption{Bayesian information criterion values of $4$ different analytic models of the splashback mass function 
versus redshifts for the two different cosmologies. }
\label{fig:bic}
\end{center}
\end{figure}

We have also modified the \st, \ca and \tn\, formulae in the same manner to accommodate the splashback mass function and compared each of them 
with the numerical results to determine their best-fit parameters. Applying the BIC method to the four analytic formulae, we have found that among 
them the numerically obtained splashback mass functions prefer most the modified \st\, and \mr\, formulae at $z\le 0.3$ and $0.3<z\le 3$, respectively 
(Figure \ref{fig:bic}).  With the help of the AIC method, it has been also shown that the numerical results prefer the modified \mr\, formula to the universal 
formula of \citet{die20a} at $0\le z < 3$.

\begin{figure}
\begin{center}
\includegraphics[scale=0.7]{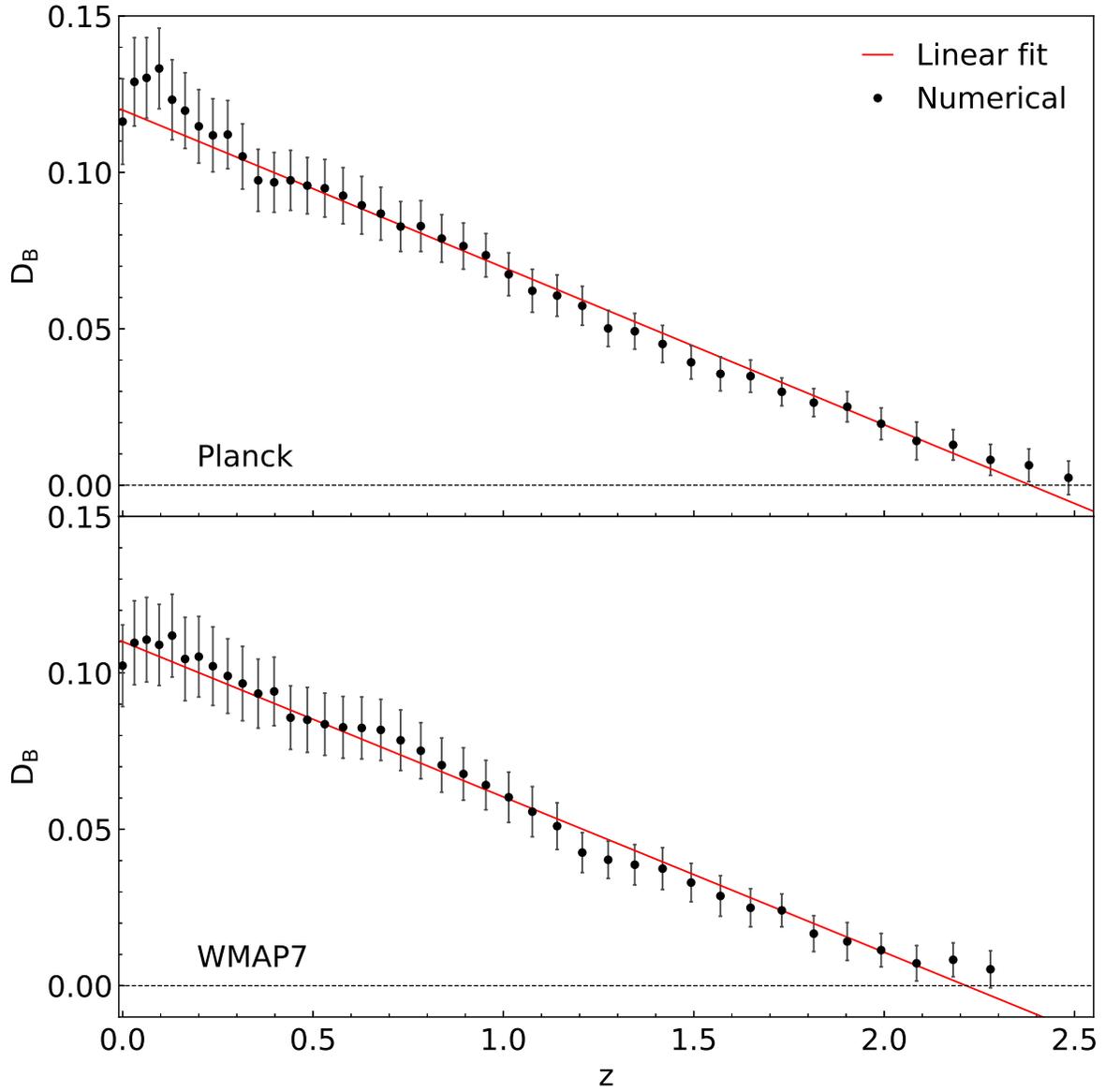}
\caption{Best-fit values of the diffusive coefficient, $D_{B}$, versus redshifts (black filled circles) and its linear fit 
given in Equation (\ref{eqn:Db_z}) in the redshift range of $0\le z\le 2.5$ }
\label{fig:Db_z}
\end{center}
\end{figure}

\begin{figure}
\begin{center}
\includegraphics[scale=0.7]{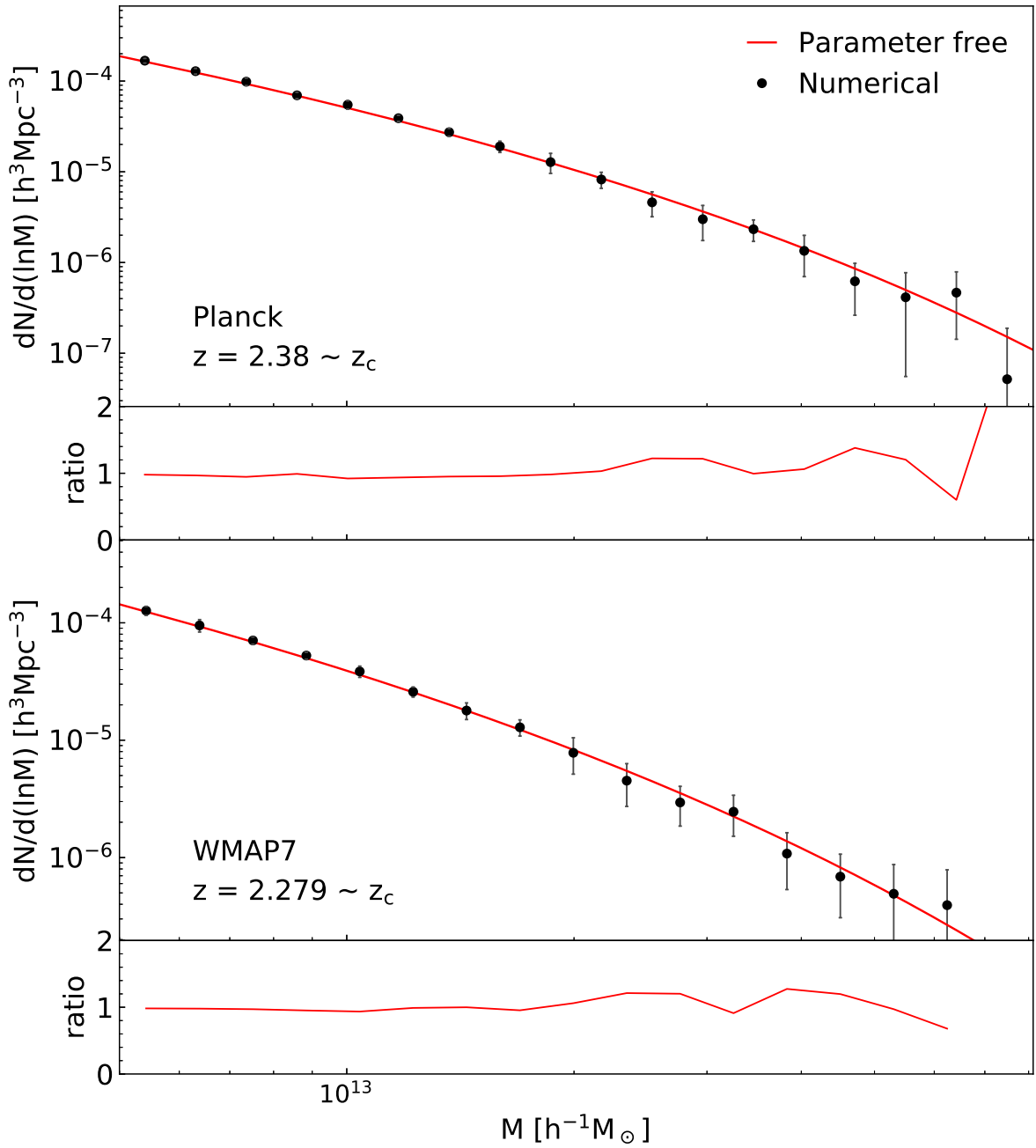}
\caption{Numerically obtained splashback mass functions (black filled circles) with the Jackknife errors 
and the parameter-free analytic formula given in Equation (\ref{eqn:para_free}) (red solid lines) at $z=z_{c}$  
for the two cosmologies.}
\label{fig:para_free}
\end{center}
\end{figure}

Despite that the modified \mr\ formula is not so universal as that of \citet{die20a} and that it turns out to be not preferred most by the BIC method 
in the redshift range of $0\le z \le 0.3$, we believe that it has a couple of advantages over the previous formulae. 
First, since it is a physical model derived analytically from the generalized excursion set theory, taking into proper account of the merit of 
identifying the DM halos with their splashback boundaries, it gives us a much better insight into the underlying connection between the initial conditions 
and the halo abundance. 
Second of all, its single parameter, the diffusion coefficient, $D_{B}$, has turned out to vary almost linearly with redshift, unlike the cases 
of the other formulae whose multiple parameters show almost desultory fluctuations with redshifts. Using this linear relation of $D_{B}(z)$, 
we have determined the critical redshift, $z_{c}$, at which $D_{B}$ vanishes (Figure \ref{fig:Db_z}). A crucial implication of this result is that 
the splashback mass functions at $z\ge z_{c}$ can be described by the purely analytical parameter-free formula (Figure \ref{fig:para_free}). 
Besides, it has turned out that the Planck and the WMAP7 cosmologies yield different values of $z_{c}$, which implies that $z_{c}$ defined as 
$D_{B}(z_{c})=0$ might be useful as a complementary probe of $\Omega_{m}$ (Table \ref{tab:sim}). 

It should be worth discussing here why the diffusion coefficient of the modified \mr\ formula exhibits a linear decrement with redshift
and why the value of the critical redshift $z_{c}$ depends sensitively on $\Omega_{m}$.  One of the primary factors that make the halo identification 
ambiguous is the tidal disturbance from the surrounding cosmic web \citep{MR10a,MR10b}. At higher redshifts, $z\sim z_{c}$, when the tidal field is 
still at its quasi-linear stage, the DM halos can be identified unambiguously by the splashback boundary. 
At lower redshifts, $z<z_{c}$, however, even the splashback boundaries would be disturbed by the more nonlinearly evolved 
tidal field \citep{web96,LS10}, which leads to a higher degree of stochasticity of the absorbing barrier and accordingly to a larger value of $D_{B}$. 

The above logic also allows us to understand why $z_{c}$ has a larger value in the Planck cosmology than in the WMAP7 cosmology.  
In the Planck cosmology with a higher value of $\Omega_{m}$ a more rapidly evolving tidal fields would disturb the halo identification at earlier times, 
which in turn deviates $D_{B}$ from zero at higher redshifts than in the WMAP7 cosmology. 
Yet, we have no clue to understanding why the decrease of $D_{B}$ follows a linear scaling with $z$. A related important question is 
whether or not this linear scaling of $D_{B}(z)$ is universal throughout the background cosmologies including the non-standard ones. If it turns out to be 
the case, then the critical density, $z_{c}$, would become a powerful probe not only of $\Omega_{m}$ but also of the background cosmology. 
Our future work will be in the direction of physically modeling $D_{B}(z)$, extending our analysis to the non-standard cosmologies and 
testing the power of $z_{c}$ as a probe of cosmology.  We hope to report the results elsewhere in the near future.

\acknowledgments

We are very grateful to our referee, B. Diemer, whose useful comments have helped us significantly improve the original manuscript. 
We acknowledge the support by Basic Science Research Program through the National Research Foundation (NRF) of 
Korea funded by the Ministry of Education (No.2019R1A2C1083855).

\clearpage
\clearpage
\clearpage
\clearpage
\clearpage
\clearpage
\clearpage
\clearpage
\clearpage
\begin{deluxetable}{cccccc}
\tablewidth{0pt}
\setlength{\tabcolsep}{5mm}
\tablecaption{Critical redshift at which the splashback mass function becomes parameter free.}
\tablehead{Cosmology &  $m_{p}$ & $\Omega_m$ & $A_{D}$ & $z_c$ \\
 & $(10^{10}\,h^{-1}\,M_{\odot})$ & & & }
\startdata
Planck &  $1.0$ & $0.32$ & $0.0503\pm 0.001$ & $2.38\pm 0.02$ \\
WMAP7 & $0.87$ & $0.27$ & $0.0497\pm 0.001$ & $2.22\pm 0.02$
\enddata
\label{tab:sim}
\end{deluxetable}


\clearpage
\begin{thebibliography}{}
\bibitem[Adhikari et al.(2014)]{adh-etal14} 
Adhikari, S., Dalal, N., \& Chamberlain, R.~T.\ 2014, JCAP, 2014, 019
\bibitem[Cavanaugh \& Neath (2019)]{aic} 
Cavanaugh, J.~E. \& Neath, A.~A.\  2019, WIREs Computational Statistics, 11, 1460, doi:10.1002/wics.1460 
\bibitem[Behroozi et al.(2013)]{rockstar} 
Behroozi, P.~S., Wechsler, R.~H., \& Wu, H.-Y.\ 2013, \apj, 762, 109
\bibitem[Bernardeau(1994)]{ber94} 
Bernardeau, F.\ 1994, \apj, 427, 51
\bibitem[Bertschinger(1985)]{ber85} 
Bertschinger, E.\ 1985, \apjs, 58, 39
\bibitem[Bond et al.(1991)]{bon-etal91} 
Bond, J.~R., Cole, S., Efstathiou, G., et al.\ 1991, \apj, 379, 440
\bibitem[Bond et al.(1996)]{web96} 
Bond, J.~R., Kofman, L., \& Pogosyan, D.\ 1996, \nat, 380, 603
\bibitem[Corasaniti \& Achitouv(2011a)]{CA11a} 
Corasaniti, P.~S. \& Achitouv, I.\ 2011a, \prl, 106, 241302
\bibitem[Corasaniti, \& Achitouv(2011b)]{CA11b} 
Corasaniti, P.~S. \& Achitouv, I.\ 2011b, \prd, 84, 023009
\bibitem[Chiueh \& Lee(2001)]{CL01} 
Chiueh, T. \& Lee, J.\ 2001, \apj, 555, 83
\bibitem[Diemer et al.(2013)]{die-etal13} 
Diemer, B., More, S., \& Kravtsov, A.~V.\ 2013, \apj, 766, 25
\bibitem[Diemer \& Kravtsov(2014)]{DK14} 
Diemer, B. \& Kravtsov, A.~V.\ 2014, \apj, 789, 1
\bibitem[Diemer \& Kravtsov(2015)]{DK15} 
Diemer, B. \& Kravtsov, A.~V.\ 2015, \apj, 799, 108
\bibitem[Diemer(2017)]{sparta17} 
Diemer, B.\ 2017, \apjs, 231, 5
\bibitem[Diemer et al.(2017)]{die-etal17} 
Diemer, B., Mansfield, P., Kravtsov, A.~V., et al.\ 2017, \apj, 843, 140
\bibitem[Diemer(2020a)]{die20a} 
Diemer, B.\ 2020a, \apj, 903, 87
\bibitem[Diemer(2020b)]{die20b} 
Diemer, B.\ 2020b, \apjs, 251, 17
\bibitem[Diemer(2020c)]{die20c} 
Diemer, B.\ 2020c, arXiv:2007.10992
\bibitem[Eke et al.(1996)]{eke-etal96} 
Eke, V.~R., Cole, S., \& Frenk, C.~S.\ 1996, \mnras, 282, 263
\bibitem[Fillmore \& Goldreich(1984)]{FG84} 
Fillmore, J.~A. \& Goldreich, P.\ 1984, \apj, 281, 1
\bibitem[Garc{\'\i}a et al.(2021)]{gar-etal21} 
Garc{\'\i}a, R., Rozo, E., Becker, M.~R., et al.\ 2021, \mnras, 505, 1195
\bibitem[Jenkins et al.(2001)]{jen-etal01} 
Jenkins, A., Frenk, C.~S., White, S.~D.~M., et al.\ 2001, \mnras, 321, 372
\bibitem[Kitayama \& Suto(1996)]{KS96} 
Kitayama, T. \& Suto, Y.\ 1996, \apj, 469, 480
\bibitem[Komatsu et al.(2011)]{wmap7} 
Komatsu, E., Smith, K.~M., Dunkley, J., et al.\ 2011, \apjs, 192, 18
\bibitem[Lahav et al.(1991)]{lah-etal91} 
Lahav, O., Lilje, P.~B., Primack, J.~R., et al.\ 1991, \mnras, 251, 128
\bibitem[Lee \& Springel(2010)]{LS10} 
Lee, J. \& Springel, V.\ 2010, JCAP, 2010, 031
\bibitem[Lewis et al.(2000)]{camb} 
Lewis, A., Challinor, A., \& Lasenby, A.\ 2000, \apj, 538, 473
\bibitem[Maggiore \& Riotto(2010a)]{MR10a}
Maggiore, M., \& Riotto, A. 2010a, \apj, 711, 907
\bibitem[Maggiore \& Riotto(2010b)]{MR10b}
Maggiore, M., \& Riotto, A. 2010b, \apj, 717, 515
\bibitem[More et al.(2015)]{mor-etal15} 
More, S., Diemer, B., \& Kravtsov, A.~V.\ 2015, \apj, 810, 36
\bibitem[Planck Collaboration et al.(2014)]{planck14} 
Planck Collaboration, Ade, P.~A.~R., Aghanim, N., et al.\ 2014, \aap, 571, A16
\bibitem[Press \& Schechter(1974)]{PS74} 
Press, W.~H., \& Schechter, P.\ 1974, \apj, 187, 425
\bibitem[Robertson et al.(2009)]{rob-etal09} 
Robertson, B.~E., Kravtsov, A.~V., Tinker, J., et al.\ 2009, \apj, 696, 636
\bibitem[Ryu \& Lee(2020a)]{RL20a} 
Ryu, S. \& Lee, J.\ 2020a, \apj, 889, 62. 
\bibitem[Ryu \& Lee(2020b)]{RL20b} 
Ryu, S. \& Lee, J.\ 2020b, \apj, 894, 65.
\bibitem[Ryu et al.(2020)]{ryu-etal20} 
Ryu, S., Lee, J., \& Baldi, M.\ 2020, \apj, 904, 93 
\bibitem[Schwarz(1978)]{bic}
 Schwarz, G.~E.\ 1978, Annals of Statistics, 6, 461
\bibitem[Shapiro et al.(1999)]{sha-etal99} 
Shapiro, P.~R., Iliev, I.~T., \& Raga, A.~C.\ 1999, \mnras, 307, 203
\bibitem[Sheth \& Tormen(1999)]{ST99} 
Sheth, R.~K. \& Tormen, G.\ 1999, \mnras, 308, 119 
\bibitem[Sheth et al.(2001)]{SMT01} 
Sheth, R.~K., Mo, H.~J., \& Tormen, G.\ 2001, \mnras, 323, 1
\bibitem[Sheth \& Tormen(2002)]{ST02} 
Sheth, R.~K., \& Tormen, G.\ 2002, \mnras, 329, 61
\bibitem[Springel(2005)]{gadget2} 
Springel, V.\ 2005, \mnras, 364, 1105
\bibitem[Tinker et al.(2008)]{tin-etal08} 
Tinker, J., Kravtsov, A.~V., Klypin, A., et al.\ 2008, \apj, 688, 709
\bibitem[Wang et al.(2009)]{wan-etal09} 
Wang, H., Mo, H.~J., \& Jing, Y.~P.\ 2009, \mnras, 396, 2249
\bibitem[Wetzel et al.(2014)]{wet-etal14} 
Wetzel, A.~R., Tinker, J.~L., Conroy, C., et al.\ 2014, \mnras, 439, 2687
\end{thebibliography}
\end{document}